\documentstyle[preprint,eqsecnum,aps,epsfig]{revtex}

\def\Journal#1#2#3#4{{#1} {\bf #2}, #3 (#4)}

\def\PRD{{\em Phys. Rev.} D}

\def\etal{{\em et al.}}
\def\vs{{\em vs.}}

\def\be{\begin{equation}}
\def\ee{\end{equation}}
\def\bea{\begin{eqnarray}}
\def\eea{\end{eqnarray}}

\def\gevcc{GeV/c$^2$}                   
\def\gevc{GeV/c}                        

\def\etmiss
{\mbox{${\hbox{$E$\kern-0.6em\lower-.1ex\hbox{/}}}_T$}}
\def\Zone{\ifmmode{\widetilde{Z_1}}\else{$\widetilde{Z_1}$}\fi}

\def\mets{{\mbox{$E\kern-0.57em\raise0.19ex\hbox{/}_{T}$}}}
\def\met{{\mbox{$E\kern-0.57em\raise0.19ex\hbox{/}_{T}$}}\ }
\def\metc{{\mbox{$E\kern-0.57em\raise0.19ex\hbox{/}_{T}^{\it cal}$}}\ }

\newcommand{\addfig}[2]{\centerline{\psfig{figure=#1,height=#2}}}
\newcommand{\x}{{\bf x}}

\begin{document}

\renewcommand{\baselinestretch}{1.0}

\preprint{
    \vbox{
            \rightline{Fermilab-Pub-00/006}\break
            \rightline{FSU-HEP-2000-0101}\break
    }
}

\title{Strategy for discovering a low-mass Higgs boson \\at the Fermilab Tevatron}

\author{Pushpalatha C. Bhat$^a$,
Russell Gilmartin$^b$, Harrison B. Prosper$^b$}

\address{$^a$Fermi National Accelerator Laboratory
\thanks{Operated by
Universities Research Association under contract to the U.S.
Department of Energy.}, Batavia, IL 60510, USA}

\address{$^b$Department of Physics, Florida State University,
Tallahassee, FL 32306}

\date{\today}

\maketitle

\begin{abstract}

We have studied the potential of the CDF and D\O\ experiments to
discover a low-mass Standard Model Higgs boson, during Run II, via
the processes $p\bar{p} \rightarrow WH \rightarrow \ell\nu
b\bar{b}$, $p\bar{p} \rightarrow ZH \rightarrow$
$\ell^{+}\ell^{-}b\bar{b}$ and $p\bar{p} \rightarrow ZH
\rightarrow \nu \bar{\nu} b\bar{b}$.  We show that a multivariate
analysis using neural networks, that exploits all the information 
contained within a set
of event variables, leads to a significant reduction, with respect
to {\em any} equivalent conventional analysis, in the integrated
luminosity required to find a Standard Model Higgs boson in the
mass range 90 \gevcc\ $ < M_H < $ 130 \gevcc. The luminosity
reduction is sufficient to bring the discovery of the Higgs boson
within reach of the Fermilab Tevatron experiments, given the anticipated
integrated luminosities of Run II, whose scope has recently been
expanded. 

\end{abstract}


\medskip

\pagebreak

\section{Introduction}

The success of the Standard Model (SM) of particle physics,
which provides an accurate
description of almost all particle phenomena observed so 
far\cite{smreview,altarelli,topreview}, 
has been spectacular. However, one crucial aspect
of it remains mysterious: the fundamental mechanism that
underlies electro-weak symmetry breaking (EWSB) and the origin of
fermion mass. Elucidating the nature of EWSB is the next major
challenge of particle physics and will be the focus of upcoming
experiments at the Fermilab Tevatron and the CERN Large Hadron Collider (LHC)
 during the early years of
the twenty-first century.

In many theories, EWSB occurs through the interaction of one or
more doublets of scalar (Higgs) fields with the initially massless
fields of the theory.  An important goal over the next decade is
to determine whether or not, in broad outline, this picture of
EWSB is correct. In the Standard Model there is a single scalar
doublet. The EWSB endows the weak bosons ($W^{\pm}, Z$) with
masses and gives rise to a single physical neutral scalar particle
called the Higgs boson ($H_{SM}$). In minimal supersymmetric
(SUSY) extensions of the SM, two Higgs doublets are required
resulting in five physical Higgs bosons: two neutral CP-even
scalars ($h,H$), a neutral CP-odd pseudo-scalar ($A$) and two charged
scalars ($H^{\pm}$). Non-minimal SUSY theories generally posit
more than two scalar doublets.

\begin{figure}
\centerline{\epsfig{file=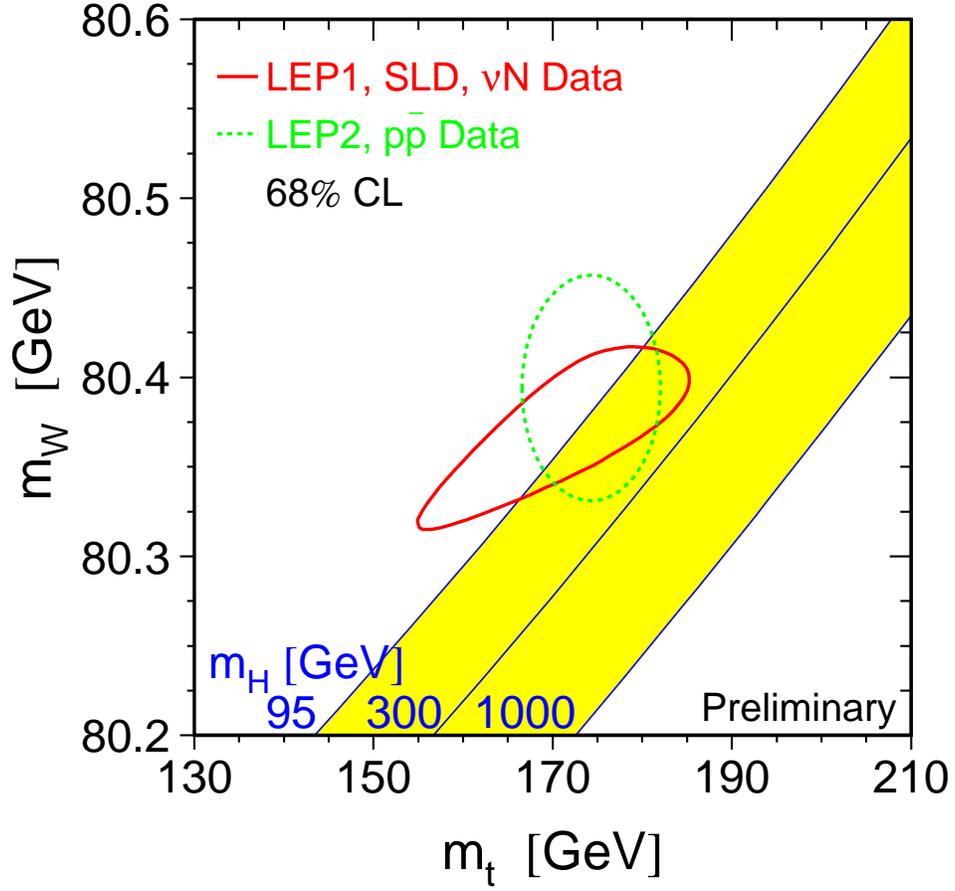,width=5in}}
\vspace{0.3in}
\caption[]{
The correlation between the $W$ boson mass and the top quark mass as
predicted by the standard model, for various possible values of the 
Higgs boson mass. (Each line corresponds to the mass value shown.)
Also shown are the 68\% CL contours from direct (dashed contour)
and indirect (solid contour) measurements of the $W$ boson and top quark mass.
From Ref.~\cite{lepewwg}.}
\label{fig:mtmw}
\end{figure}

Given this picture of EWSB, 
the direct and indirect measurements of the top
quark and $W$ boson masses
constrain the mass of the SM
Higgs boson ($M_{H_{SM}}$), as indicated in Fig~\ref{fig:mtmw}.
A global fit to all electroweak precision data,
including the top quark mass, gives a central value of
$M_{H_{SM}} = 107_{-45}^{+67}$ GeV/c$^2$ and a 95\% confidence level
upper limit of 225 GeV/c$^2$\cite{smreview}.
In broad classes
of SUSY theories the mass $M_h$ of the lightest CP-even neutral
Higgs boson, $h$, is constrained to be less than 150~\gevcc
\cite{martin}. In the minimal supersymmetric SM (MSSM), the upper
bound on $M_h$ is lowered to about 130 \gevcc \cite{carena,ukreport}. 
This bound is reasonably robust
with respect to changes in the parameters of the theory.
Furthermore, in the limit of large pseudo-scalar Higgs boson mass, $M_A >>
M_Z$, where $M_Z$ is the mass of the $Z$ boson, the properties of
the lightest MSSM Higgs boson $h$ are indistinguishable from those
of the SM Higgs boson, $H_{SM}$. These intriguing indications of a
low-mass Higgs boson motivate the study of strategies that
maximize the potential for its discovery at the upgraded
Tevatron\cite{workshop}.
This paper describes a strategy that achieves
this goal.

\begin{figure}
\centerline{\epsfig{file=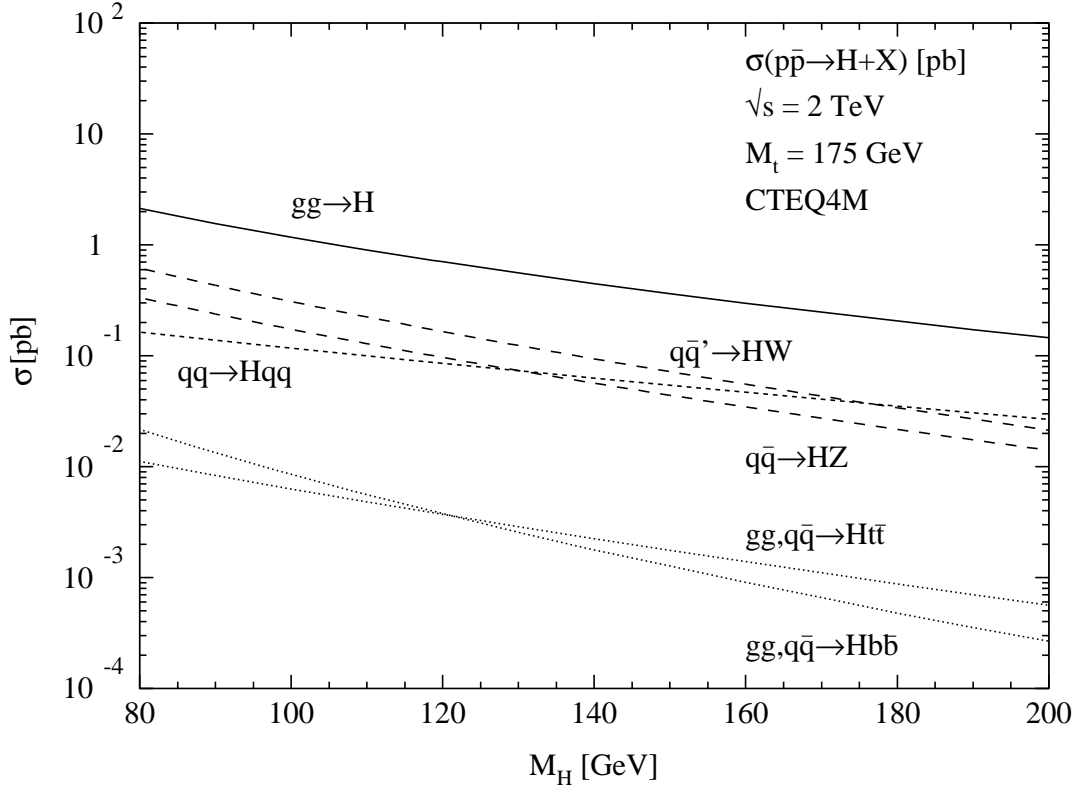,width=5in,angle=270}}
\vspace{0.3in}
\caption[]{
Cross sections for various Higgs  production processes in
$p\bar{p}$ collisions at $\sqrt{s} = 2$ TeV as a
function of Higgs boson mass. From Ref.~\cite{spira}.}
\label{fig:crsecplot}
\end{figure}

The current 95\% 
CL  lower limit on the Higgs boson mass, from the CERN $e^+e^-$ collider LEP,
is 107.9~\gevcc \cite{lepc} and is expected to reach close to 
114~\gevcc \cite{ukreport} in
the near future. We have therefore studied the mass range
90~\gevcc\ $< M_H <$ 130 \gevcc, where $H$, hereafter, denotes the SM
Higgs boson, $H_{SM}$. The cross sections for SM Higgs boson production at the 
Fermilab Tevatron are
shown in Fig~\ref{fig:crsecplot}. At $\sqrt s = 2$~TeV, the dominant
process for the production of Higgs bosons in $p\bar{p}$
collisions is $gg \rightarrow H$. The Higgs boson decays to a $b\bar{b}$
pair about 85\% 
of the time. Unfortunately, even with maximally
efficient $b$-tagging this channel is swamped by QCD di-jet
production. The more promising channels are $p\bar{p}\rightarrow
WH\rightarrow \ell\nu b\bar{b}$, $p\bar{p}\rightarrow
ZH\rightarrow$ $\ell^{+}\ell^{-}$ $b\bar{b}$ and
$p\bar{p}\rightarrow$ $ZH\rightarrow$ $\nu \bar{\nu} b\bar{b}$, which are the 
ones we have studied.
 
In $WH$ events the lepton can be lost because of deficiencies in the
detector or the event reconstruction or the lepton energy being below
the selection threshold.  For such events the {\em reconstructed}
final state would be indistinguishable from that arising from the
process $p\bar{p}\rightarrow$ $ZH\rightarrow$ $\nu \bar{\nu}
b\bar{b}$.  We have therefore studied these processes in terms of the
channels: {\em single lepton} ($\ell$ + \met + $b\bar{b}$ from 
$WH$), 
{\em di-lepton} ($\ell^{+}\ell^{-} b\bar{b}$ from $ZH$)
 and { \em missing transverse energy}
(\met+ $b\bar{b}$ from $ZH$ and  $WH$), where
$\met$ denotes the missing transverse energy from all sources,
including neutrinos.  For each of these channels, we have carried out
a comparative study of multivariate and conventional analyses of these
channels in which we compare signal significance and the integrated
luminosity needed for discovery.

The paper is organized as follows: In Sec.~\ref{sec:optimal}
we describe our strategy in general terms. Sections~\ref{sec:lepton},
\ref{sec:dilepton} and \ref{sec:missinget}, respectively, 
describe our analyses of the single lepton, di-lepton and 
missing transverse energy channels.
Our conclusions are given in Sec.~\ref{sec:summary}.

\section{Optimal Event Selection}
\label{sec:optimal} 
In conventional analyses a cut is
applied to each event variable, usually one variable at a time,
after a visual examination of the signal and background
distributions. Although analyses done this way are
sometimes described as
``optimized," in practice, unless the signal
and background distributions are well separated, the
traditional procedure for choosing cuts is rarely optimal
in the sense of
{\em minimizing the probability to mis-classify events.}
Since we wish to maximize the chance of discovering the Higgs
boson we need to achieve the optimal separation between signal and
background, while maximizing the signal significance. Given any
set of event variables, optimal separation can always be achieved
if one treats the variables in a fully {\em multivariate} manner.

Given a set of event variables, it is useful to construct the
discriminant function $D$ given by
\begin{equation}
D =\frac{s(\x)}{s(\x)+b(\x)},
\label{eq:D}
\end{equation}
 where $\x$ is the vector of variables that characterize the events and
 $s(\x)$ and $b(\x)$, respectively,
are the $n-$dimensional probability densities describing the
signal and background distributions. The discriminant function $D
= r/(1+r)$ is related to the {\em Bayes discriminant function}
which is proportional to the likelihood ratio $r
\equiv s(\x)/b(\x)$. Working with $D$, instead of directly with $\x$,
brings two important advantages: 1) it reduces a difficult
$n-$dimensional optimization problem to a trivial one in a single
dimension and 2) a cut on $D$ can be shown to be optimal in the
sense defined above.

There is, however, a practical difficulty in calculating the
discriminant $D$. We usually do not have analytical expressions
for the distributions $s(\x)$ and $b(\x)$. What is normally
available are large discrete sets of points $\x_i$, generated by
Monte Carlo simulations. Fortunately, however, there are several
methods available to approximate the discriminant $D$ from a set
of points $\x_i$,  the most convenient of which uses feed-forward
neural networks. Neural networks are ideal in this regard because
they approximate $D$ directly\cite{bhat,blum}.

Many neural network packages are available, any one of which can
be used to calculate $D$. We have used the JETNET
package\cite{jetnet} to train three-layer (that is, input, hidden
and output) feed-forward neural networks (NN). The training was done
using the back-propagation algorithm, with the target output for
the signal set to one and that for the background set to zero.
In this paper we
use the terms ``neural network output"
and ``discriminant" interchangeably.
However, the distinction
between the exact discriminant $D$, as
we have defined it above, and the network output,
which provides an estimate of $D$, should be borne in mind.

\section{Single Lepton Channel}
\label{sec:lepton}

We have considered final states with a high $p_T$ electron~(e) or
muon~($\mu$) and a neutrino from $W$ decay and a $b\bar{b}$ pair
from the decay of the Higgs boson. The $WH$ events were simulated using
the PYTHIA program\cite{pythia} for Higgs boson masses of $M_H$ =
90, 100, 110, 120 and 130 \gevcc. In
Table I we list the cross section $\times$
branching ratio (BR) we have used for the process
$p\bar{p}\rightarrow WH\rightarrow \ell\nu b\bar{b}$ where
$\ell=e,\mu$, $\tau$.

The processes $p\bar{p} \rightarrow Wb\bar{b}$, $p\bar{p}
\rightarrow WZ$, $p\bar{p} \rightarrow t\bar{t}$, single top
production---$p\bar{p} \rightarrow W^* \rightarrow tb$ and
$p\bar{p} \rightarrow Wg \rightarrow tqb$, which have the same
signature, $\ell \nu b \bar{b}$, as the signal, are the most
important sources of background. They have all been included in our
study.  The $Wb\bar{b}$ sample was generated using
CompHEP\cite{comphep}, a parton level Monte Carlo program based on
exact leading order (LO) matrix elements. The parton fragmentation
was done using PYTHIA. The single top, $t\bar{t}$ and $WZ$ events
were simulated using PYTHIA. To generate the s-channel process,
$W^* \rightarrow tb$, we forced the $W$ to be produced off-shell,
with $\sqrt{\hat{s}}
> m_t + m_b$, and then selected the final state in which $W \rightarrow t b$.
The cross sections used for the background processes are given in
Table I.

To model the expected response of the CDF and D\O\ Run II detectors 
at Fermilab we
used the SHW program\cite{SHW}, which provides a
fast (approximate) simulation of the trigger, tracking,
calorimeter clustering, event reconstruction and $b$-tagging.
The SHW simulation predicts a di-jet mass resolution of about 14\%
at $M_H$ = 100~\gevcc, varying only slightly over the mass
range of interest.  However, to allow for comparisons with the
other $WH$ and $ZH$  studies at the Physics at Run II SUSY/Higgs
workshop\cite{workshop}, some of which do not use SHW, we have
re-scaled the di-jet mass variables for all signal and background
events so that the resolution is 10\% 
at each Higgs boson mass.
The consensus of Run II workshop is that such a mass resolution
can be achieved, albeit with considerable effort.

In principle, multivariate methods can be applied at all stages of
an analysis. However, in practice, experimental considerations,
such as trigger thresholds and the need to restrict data to the
phase space in which the detector response is well understood,
dictate a set of loose cuts on the event variables. These cuts
define a {\em base} sample of events. In our case, the base sample
was determined by the following cuts:
\begin{itemize}
\item
the transverse momentum of the isolated lepton
$P_T^{\ell} > 15$~\gevc\
\item
the pseudo-rapidity of the lepton
$|{\eta}_{\ell}| < 2$
\item
the missing transverse energy in the event
$\met > 20$~GeV
\item
two or more jets in the event
with $E_T^{jet} > 10$~GeV and $|{\eta}_{jet}| < 2$.
\end{itemize}
Since the Higgs decays into a $b\bar{b}$ pair we impose the requirement
that two jets be $b$-tagged. This of course does
little to  reduce the dominant $Wb\bar{b}$ background,
due to the presence of
the $b\bar{b}$ pair, but it becomes powerful when the invariant
mass, $M_{b\bar{b}}$, of the $b$-tagged jets is used as an event
variable. The
di-jet mass distributions for the signal is expected to peak
at the Higgs boson mass, whereas one expects a broad distribution for the
background, with the exception of the $WZ$ background which peaks at the
$Z$ boson mass.

One of the $b$-tags was required to be {\it tight}
and the other
{\it loose}\cite{SHW}. A tight $b$-tag is defined by an algorithm that uses
the silicon vertex detector, while a loose $b$-tag is defined by
the same algorithm with looser cuts or by a soft lepton
tag\cite{SHW}. The mean double $b$-tagging efficiency in
SHW is about 45\%.

We searched for variables that discriminate between the signal
and the backgrounds and arrived at the following set:

\begin{itemize}
\item
$E_T^{b1}, E_T^{b2}$ -- transverse energies of the $b$-tagged jets
\item
$M_{b\bar{b}}$ -- invariant mass of the $b$-tagged jets
\item
$H_T$ -- sum of the transverse energies of all selected jets
\item
$E_T^\ell$ -- transverse energy of the lepton
\item
$\eta_\ell$ -- pseudo-rapidity of the lepton
\item
$\met$ -- missing transverse energy
\item
$S$ -- sphericity ($S = \frac{3}{2}(Q_1 + Q_2)$ where
$Q_i$ are the eigenvalues obtained by diagonalizing
the normalized momentum tensor $M_{ab} = \sum_i p_{ia} p_{ib}/
\sum_i|p_i|^2$ where the sums are over the final state particle
momenta and the subscripts $a$ and $b$ refer to the spatial components
of the momenta $p_i$
\item
$\Delta R(b_1,b_2)$ -- the distance, in the $(\eta, \phi)$-plane, between
the two $b$-tagged jets, where 
$\Delta R = \sqrt{\Delta\eta^2 + \Delta\phi^2}$ and $\phi$ is the
azimuthal angle

\item
$\Delta R(b_1,\ell)$ -- the $\Delta R$ distance 
between the lepton and the first $b$-tagged jet.
\end{itemize}

Most of the variables used are directly measured (reconstructed)
kinematic quantities while some are deduced variables. The choice of
$M_{b\bar{b}}$ as a discriminating variable is obvious, as discussed
earlier. The variable $H_T$ is a measure of the ``temperature'' of the
interaction; a large $H_T$ is a sign of the decay of massive objects.
For example, $WH$ events would have larger $H_T$ (increasing with $M_H$)
than the $Wb\bar{b}$ background, but smaller $H_T$ than the $t\bar{t}$
background.  The $WH$ events are also more spherical than the
$Wb\bar{b}$ events and  have larger values of sphericity.  The
$\Delta R(b,\bar{b})$ is smaller for $Wb\bar{b}$ background where the
$b$-jets come mainly from $g\rightarrow b\bar{b}$  than  in 
$WH$ events where the $b$-jets come from the heavy object decay $H\rightarrow
b\bar{b}$.

\begin{figure}
\centering
    \begin{minipage}[t]{.45\linewidth}
    \mbox{\epsfig{file=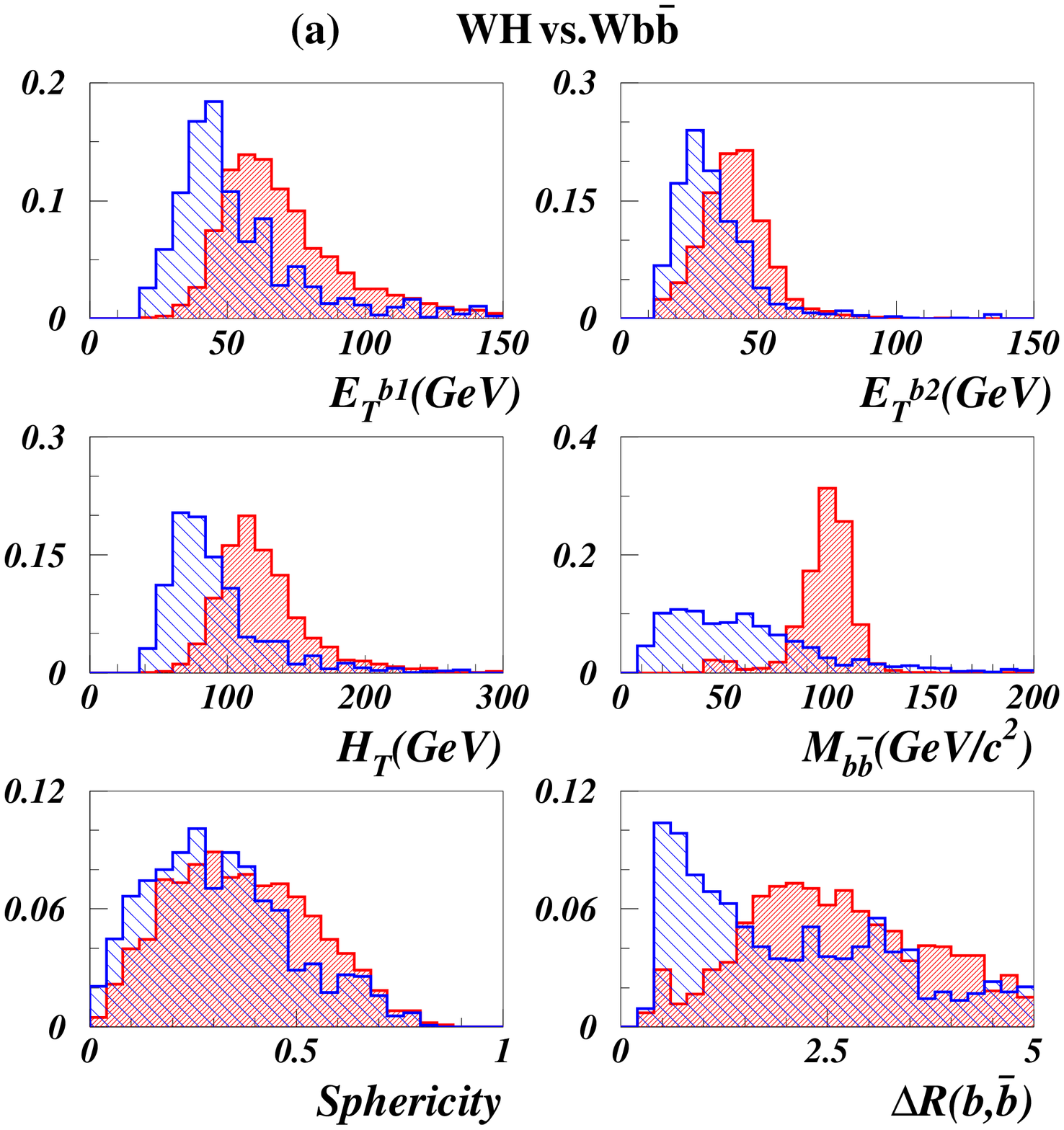, width=3.2in}}
    \end{minipage}
    \begin{minipage}[t]{.45\linewidth}
    \mbox{\epsfig{file=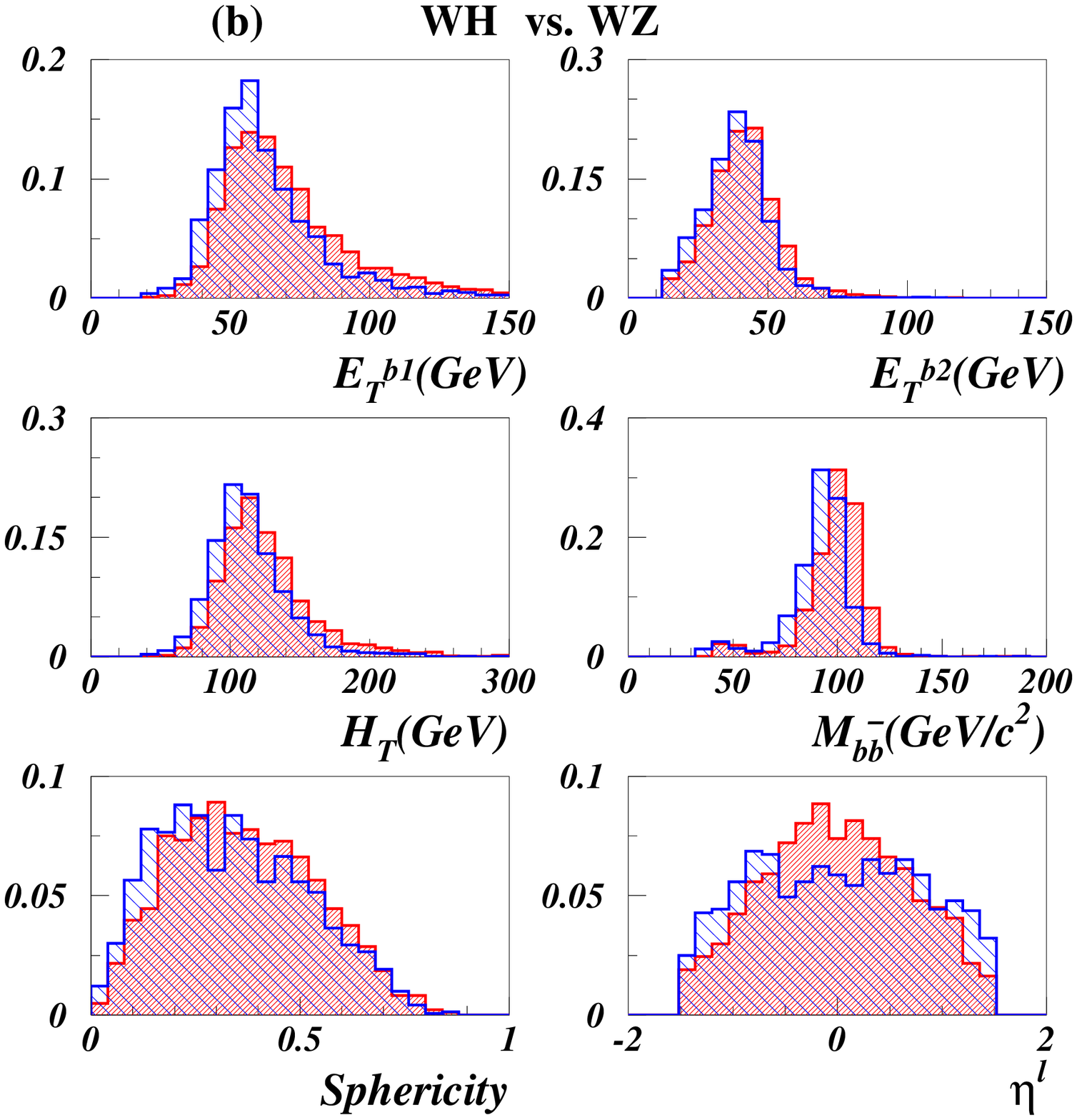, width=3.2in}}
    \end{minipage}
    \begin{minipage}[b]{.45\linewidth}
    \mbox{\epsfig{file=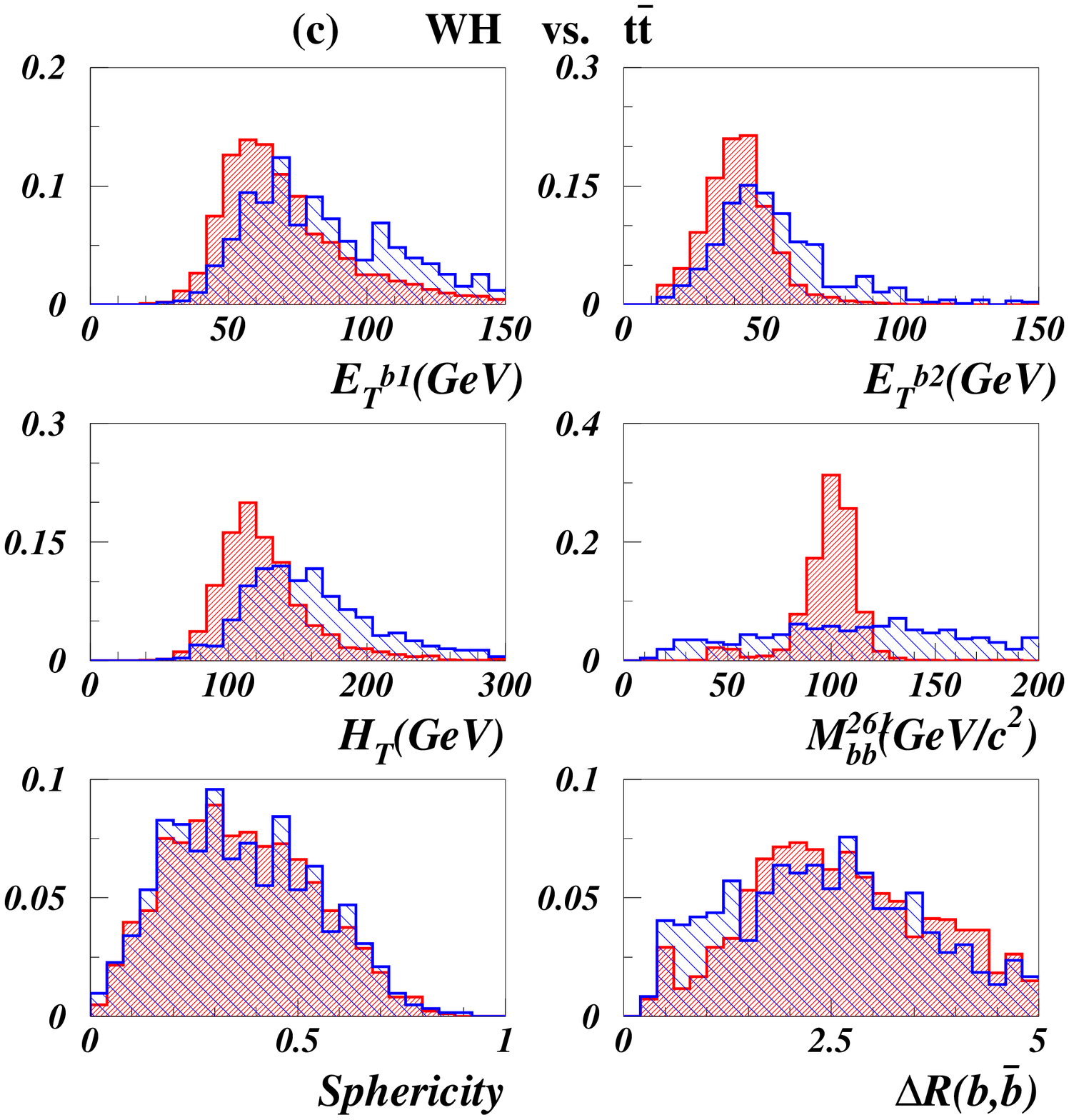, width=3.2in}}
    \end{minipage}
    \begin{minipage}[b]{.45\linewidth}
    \mbox{\epsfig{file=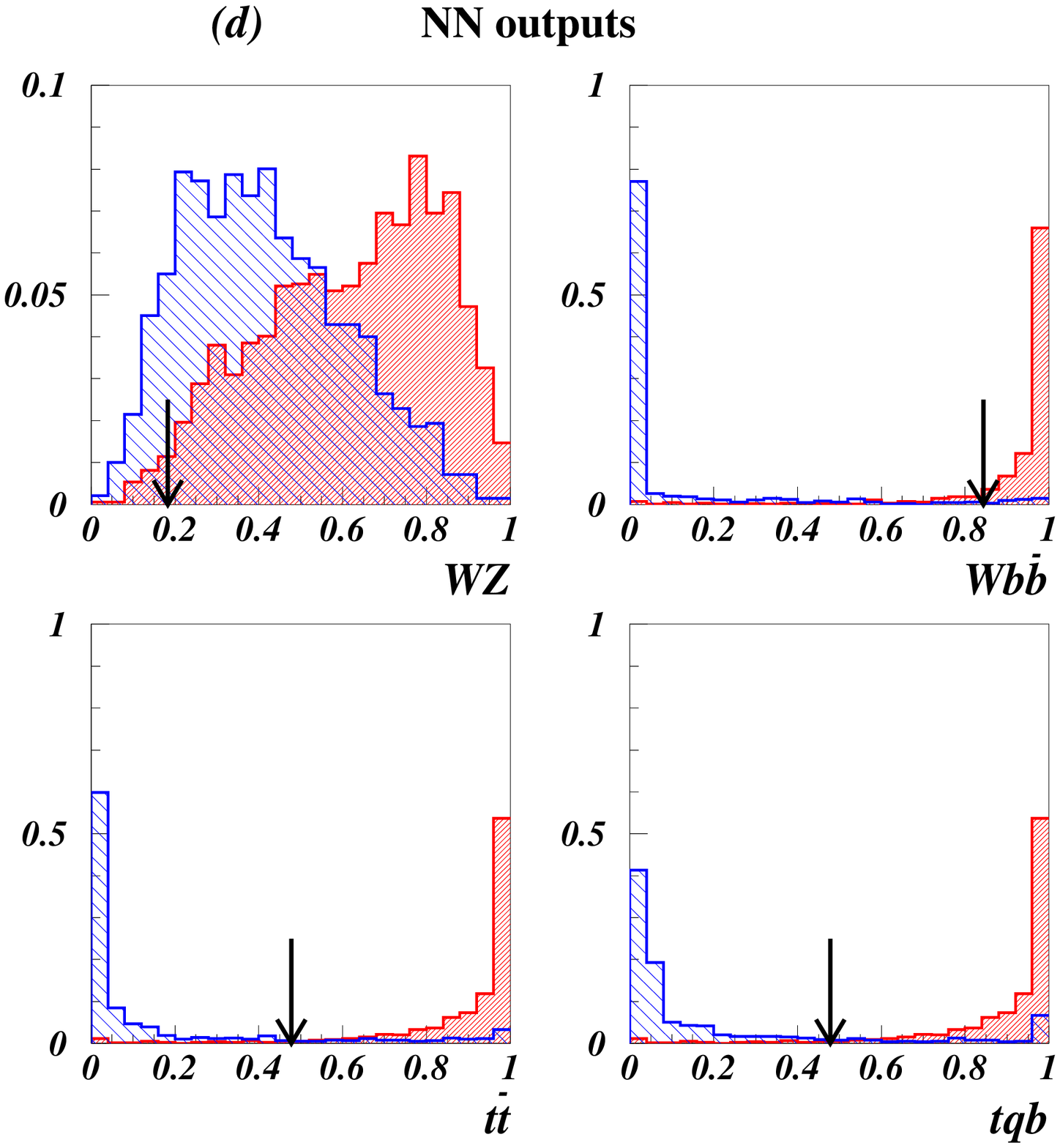, width=3.2in}}
    \end{minipage}
\vspace{0.3in}

\caption[]{ Distributions of some of the variables used in the NN
analysis for $WH$ ($M_H$=100 \gevcc) signal (heavily shaded) and
backgrounds (lightly shaded) (a) $WH$ {\em vs.} $Wb\bar{b}$, (b) $WH$ 
{\em vs.} $WZ$,
(c) $WH$ {\em vs.} $t\bar{t}$. In (d) we compare the neural network output 
distributions for signal and various backgrounds. 
The arrows indicate the cuts.}
\label{fig:wh}
\end{figure}

For each Higgs boson mass we trained three networks to discriminate
against the main
backgrounds $Wb\bar{b}$, $WZ$ and $t\bar{t}$.
The subsets of variables used to train the networks
are listed in Table II while in Fig~\ref{fig:wh}(a-c)
we show the
distributions of some
of these variables.
Each network has
7 input variables, 9 hidden nodes and
one output node.
We calcuated three discriminants $D$ for every
signal and background event and for every Higgs boson mass.
Figure~\ref{fig:wh}(d) shows the distributions of the discriminants
for signal and background calculated using
the network trained to discriminate between signal events, with  
$M_H$ = 100~\gevcc, and the specified background.
  We note that all backgrounds, with the exception of
$WZ$, are well separated from the signal. For Higgs boson masses
close to the $Z$ mass
the $WZ$ background is kinematically identical to the signal and
therefore difficult to deal with.
But for Higgs boson masses well above the $Z$ mass the discrimination
between $WH$ and
$WZ$ improves, as does that between $WH$ and the other backgrounds.
(In all figures, the signal histograms are shaded dark
while the background histograms are shaded light.) 
 The arrows  in Fig.~\ref{fig:wh}(d) indicate the 
cuts applied to the discriminants.
The cuts were chosen to maximize $S/\sqrt B$, where
$S$ and $B$ are the signal and background counts, respectively.  
The cuts to suppress the $WZ$ background vary from 0.18 to 0.80, increasing
for higher Higgs boson masses; the cuts to suppress $Wb\bar{b}$ are generally 
about 0.8, while those for top events are in the range 0.35 to 0.75.

\begin{figure}
\addfig{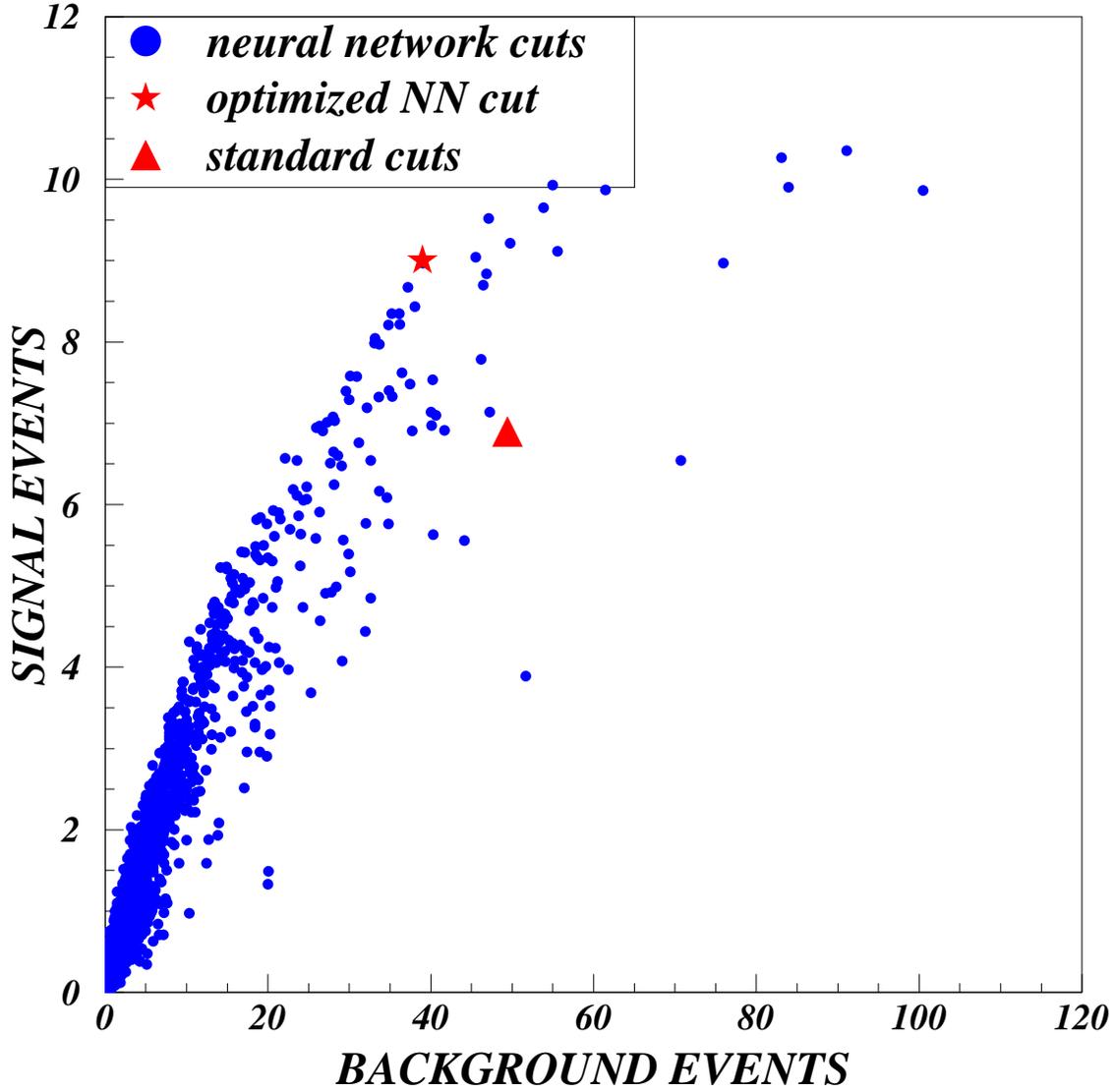}{6.5in} 
\vspace{0.3in}

\caption[]{Single lepton channel. The number of
signal events vs. number of background events for 1 fb$^{-1}$
using various combination of cuts on the three neural network
outputs. The standard cuts are optimized based on studies
done in the Higgs working group using conventional methods.
} 
\label{fig:rgsplot}
\end{figure}

At this stage it is instructive to compare the conventional and
multivariate approaches, to assess what has been gained by using
the latter approach.
In
Fig.~\ref{fig:rgsplot} we compare the signal efficiency \vs\
background efficiency (given in terms of the number of events 
for 1 fb$^{-1}$) for
an ensemble of possible cuts on the three discriminants
(using the random grid search technique\cite{rgs})
with the efficiencies  obtained using the standard cuts
defined by the Run II Higgs Workshop\cite{workshop}.
 Each dot corresponds to a particular set
of cuts on the three discriminants; 
the triangular marker indicates what is
achieved using the standard cuts, while the star indicates the results obtained
from an optimal choice of cuts (which maximizes $S/\sqrt B$)
on the three network outputs.
Table III shows results for the $WH$ channel.

\section{Di-Lepton  Channel}
\label{sec:dilepton}

For the di-lepton channel we followed a strategy similar to that
described for the single lepton channel.
The final state signature considered is: two high $P_T$
same flavor leptons ($ee$ or $\mu\mu$) from $Z$ boson decay and two
b-jets (from $H\rightarrow b\bar{b}$).

The $ZH$ events were generated using PYTHIA for
Higgs boson masses of 90, 100, 110, 120 and 130 GeV/c$^2$.
The principal backgrounds are due to $ZZ$, $Zb\bar{b}$,
single top and $t\bar{t}$ production. The $Zb\bar{b}$ background sample was
generated using CompHEP, with fragmentation done using PYTHIA, while all
other samples were generated using PYTHIA.
As before, the SHW program was used to simulate the detector response
and we assumed that two jets are $b$-tagged (one tight and one loose).
The cross sections for signal and background are shown
in Table I.
The base sample was determined by the following cuts:
\begin{itemize}
\item
$P_T^{\ell} > 10$~\gevc\

\item
$|{\eta}_{\ell}| < 2$

\item
$\met < 10$~GeV
\item
at least two jets
with $E_T^{jet} > 8$~GeV and $|{\eta}_{jet}| < 2$.
\end{itemize}

A network was trained for each Higgs boson mass
and for each of the three backgrounds with the following variables
\begin{itemize}
\item
$E_T^{b1},E_T^{b2}$
\item
$P_T$ of the two leptons
\item
$M_{b\bar{b}}$
\item
$M_{\ell\bar{\ell}}$ -- invariant mass of the leptons
\item
$H_T$
\item
$\Delta R(b_1,\ell)$ between the first lepton and the first $b$-tagged jet.
\end{itemize}

\begin{figure}
\centering
    \begin{minipage}[t]{.45\linewidth}
    \mbox{\epsfig{file=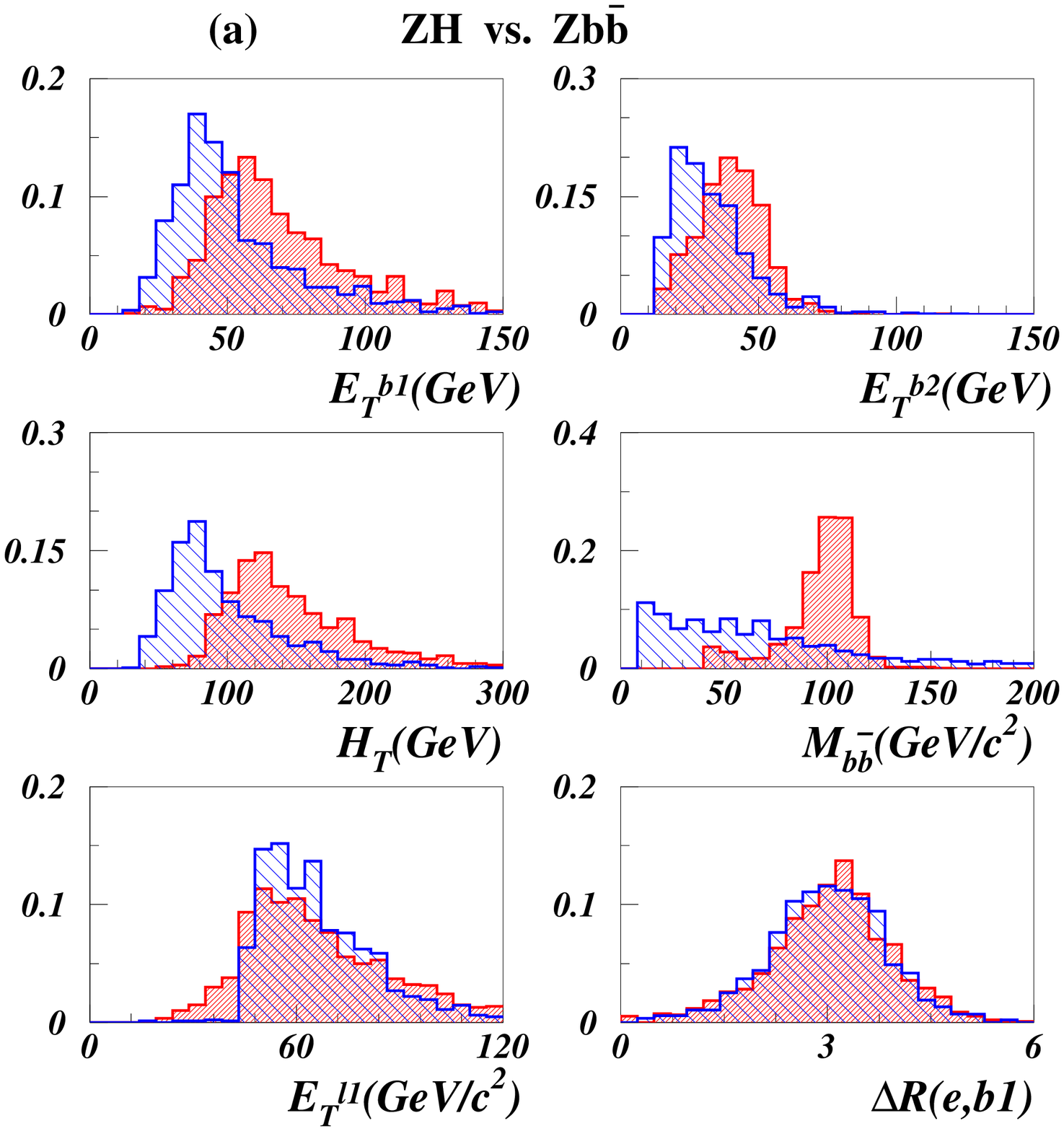, width=3.2in}}
    \end{minipage}
    \begin{minipage}[t]{.45\linewidth}
    \mbox{\epsfig{file=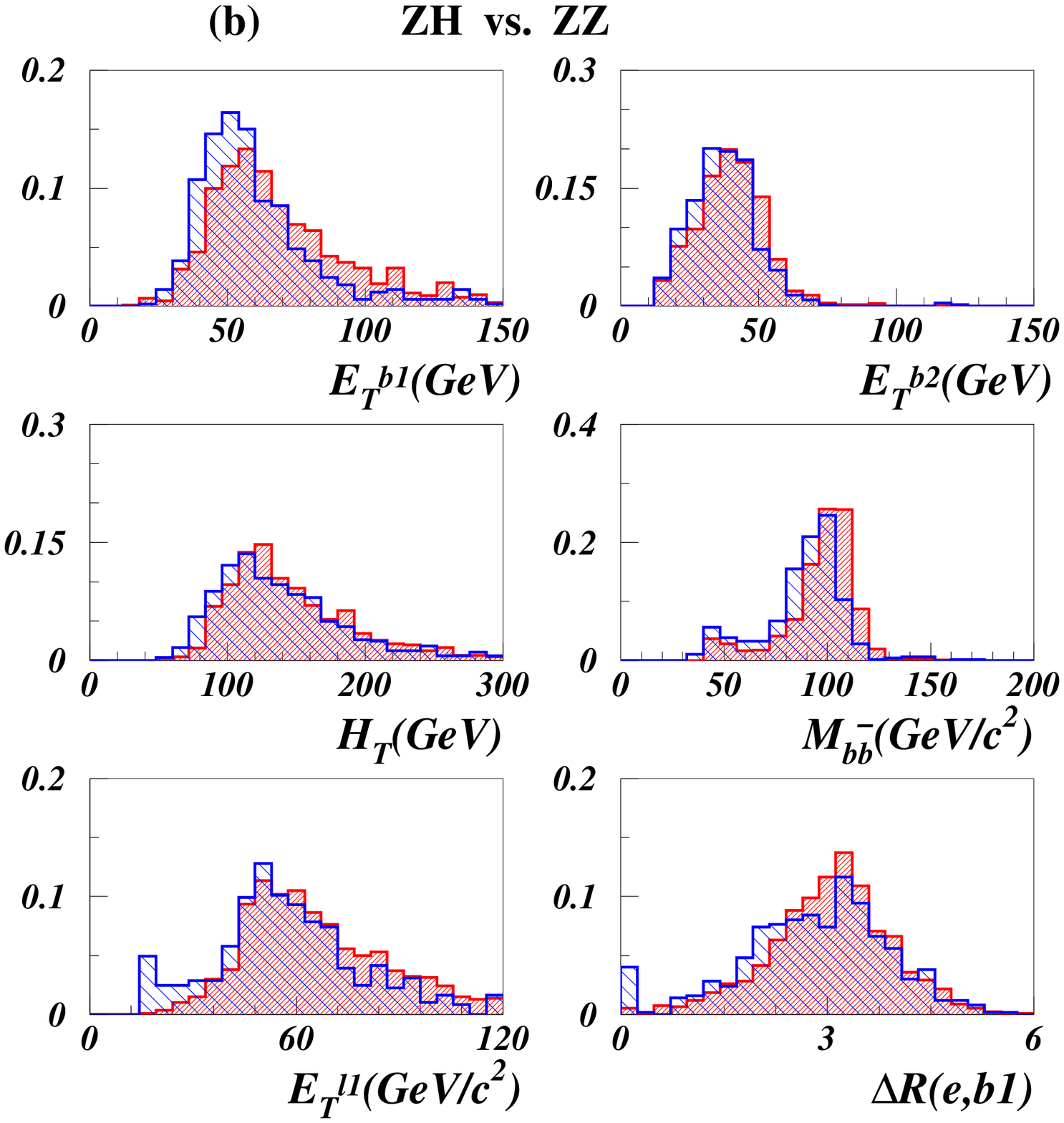, width=3.2in}}
    \end{minipage}
    \begin{minipage}[b]{.45\linewidth}
    \mbox{\epsfig{file=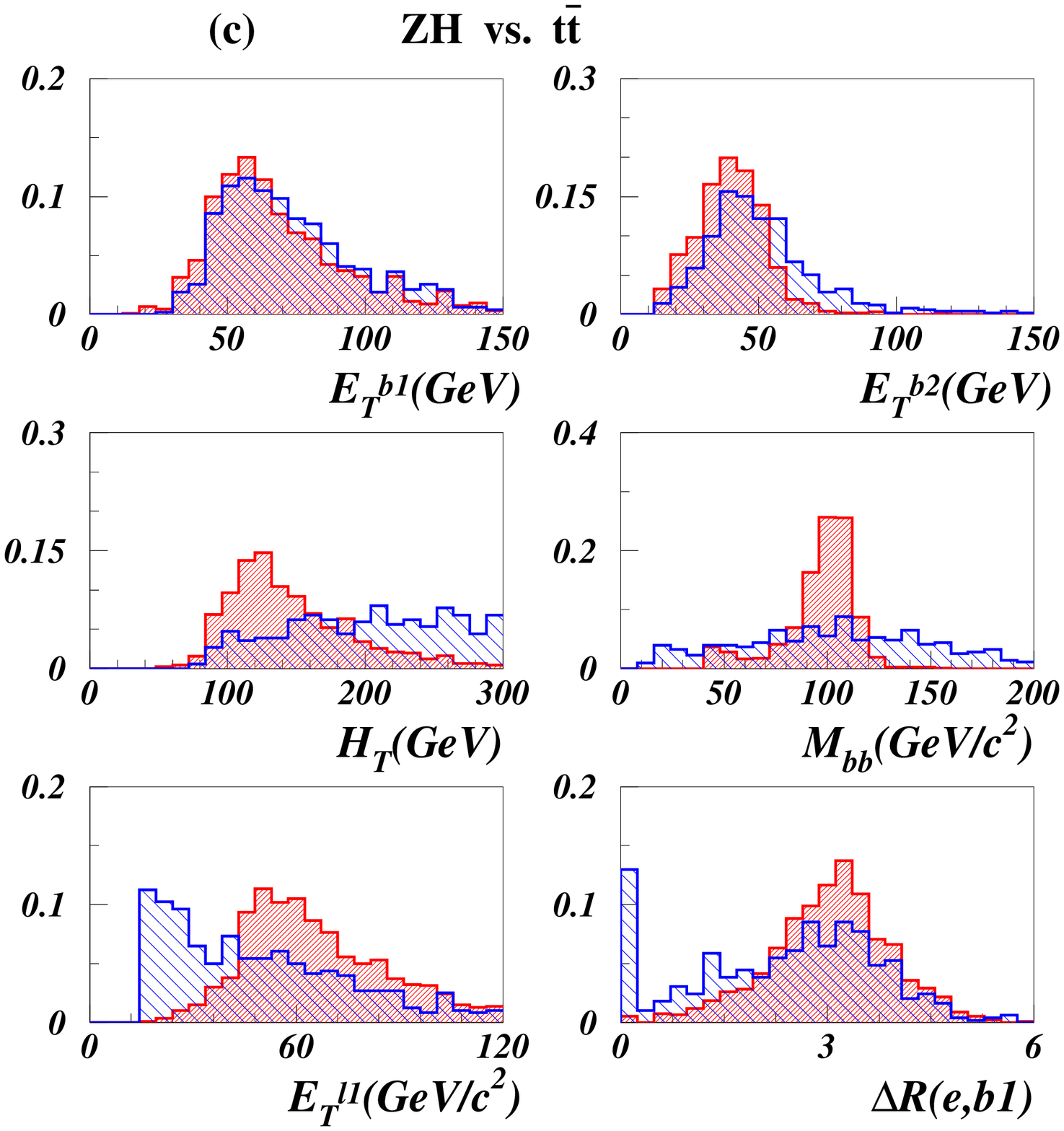, width=3.2in}}
    \end{minipage}
    \begin{minipage}[b]{.45\linewidth}
    \mbox{\epsfig{file=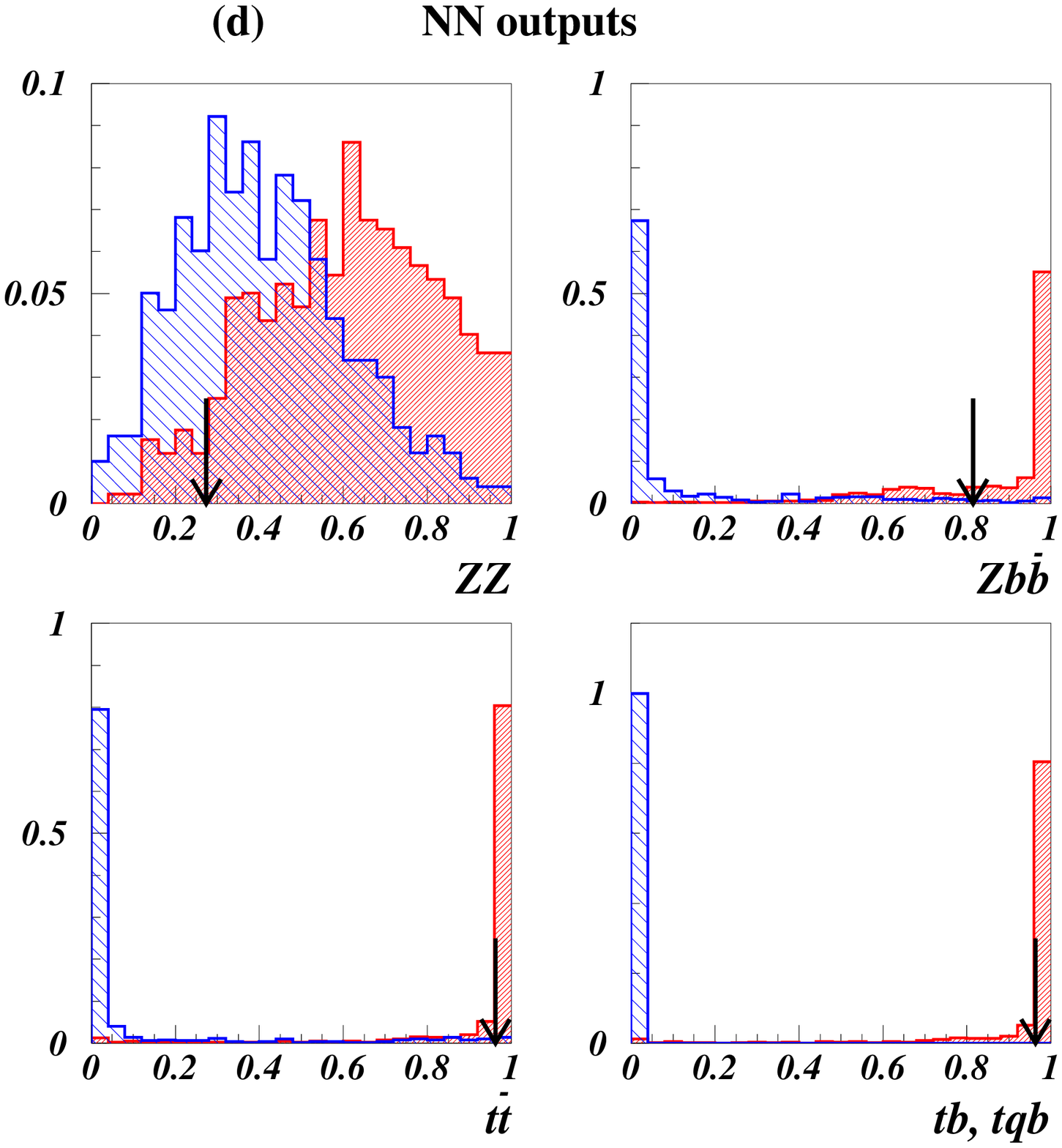, width=3.2in}}
    \end{minipage}
\vspace{0.3in}

\caption[]{Di-lepton channel.
Distributions of variables used in training the neural networks
for signal (with $M_H=100$~\gevcc) and different backgrounds and
the results of the trained networks. (a) Signal {\em vs.} $Zb\bar{b}$
background; (b) signal {\em vs.} $ZZ$ background; (c) signal {\em vs.}
$t\bar{t}$ background and (d) distributions of neural network outputs
for networks trained using signal {\em  vs.} the backgrounds 
$ZZ$, $Zb\bar{b}$ and $t\bar{t}$. The signal histograms are heavily shaded. 
The arrows indicate the cuts.
}

\label{fig:zhll}
\end{figure}

Distributions of these variables, as well as 
those of the network output, are shown
in Fig~\ref{fig:zhll}(a-d). 
The signal distributions are for $M_H$=100 GeV/c$^2$.
Our results after applying cuts on the three network outputs,
for the di-lepton channels are summarized in
Table IV.

\section{Missing Transverse Energy  Channel}
\label{sec:missinget}
This channel has contributions from both $ZH \rightarrow \nu\bar{\nu} b\bar{b}$
and $WH \rightarrow (\ell) \nu b\bar{b}$ where $(\ell)$ 
denotes the lepton that is
lost. The event generation and detector simulation were carried out as 
described in the single lepton and di-lepton channel studies. The
base sample was defined by the cuts

\begin{itemize}
\item
$|{\eta}_{\ell}| < 2$
\item
$\met > 10$~\gevc\
\item
no isolated lepton with
$P_T^{\ell} > 10$~\gevc\
\item
$E_T^{jet3} < 30$~GeV 
\item
at least two jets
with $E_T^{jet} > 8$~GeV and $|{\eta}_{jet}| < 2$.
\end{itemize}

\noindent The three networks were trained 
with $ZH\rightarrow \nu\bar{\nu}b\bar{b}$ events as signal and
$Zb\bar{b}$, $ZZ$ and $t\bar{t}$ as the three backgrounds, respectively.
The same networks were used to evaluate contributions from $WH$ 
and the relevant backgrounds.
We used the following variables to train the networks: 

\begin{itemize}
\item
$E_T^{b1},E_T^{b2}$
\item
$M_{b\bar{b}}$
\item
$H_T$
\item
$\met$
\item
$S$
\item
${\cal C}$ -- centrality ($\sum_{jets} E_T/\sum_{jets} E$,
 with $E_T^{jet} > 15$~GeV)

\item
$\frac{\met}{\sqrt{E^{b1}_T}}$
\item
minimum $\Delta\phi(jet,\met)$.
\end{itemize}

\begin{figure}
\centering
    \begin{minipage}[t]{.45\linewidth}
    \mbox{\epsfig{file=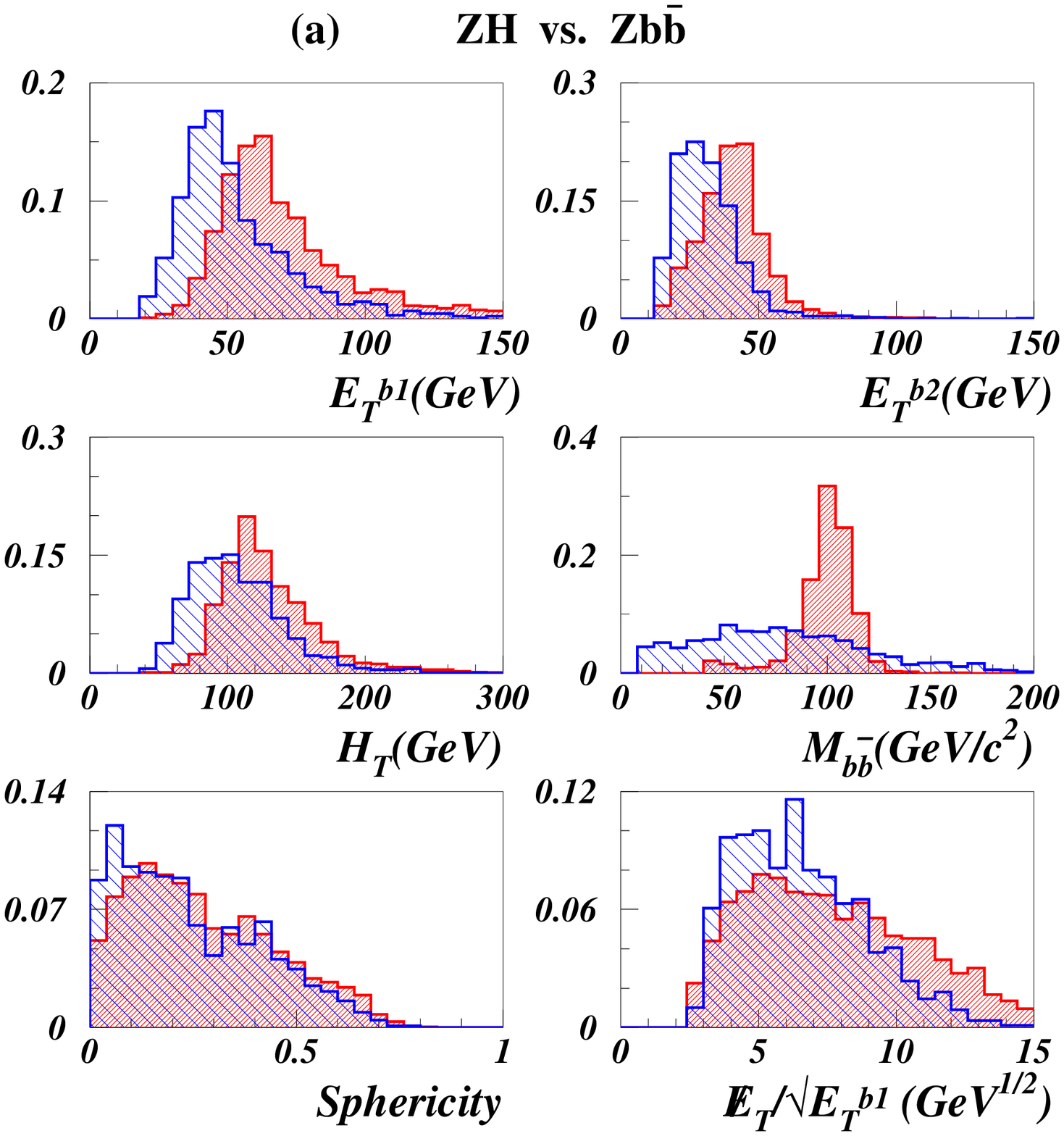, width=3.2in}}
    \end{minipage}
    \begin{minipage}[t]{.45\linewidth}
    \mbox{\epsfig{file=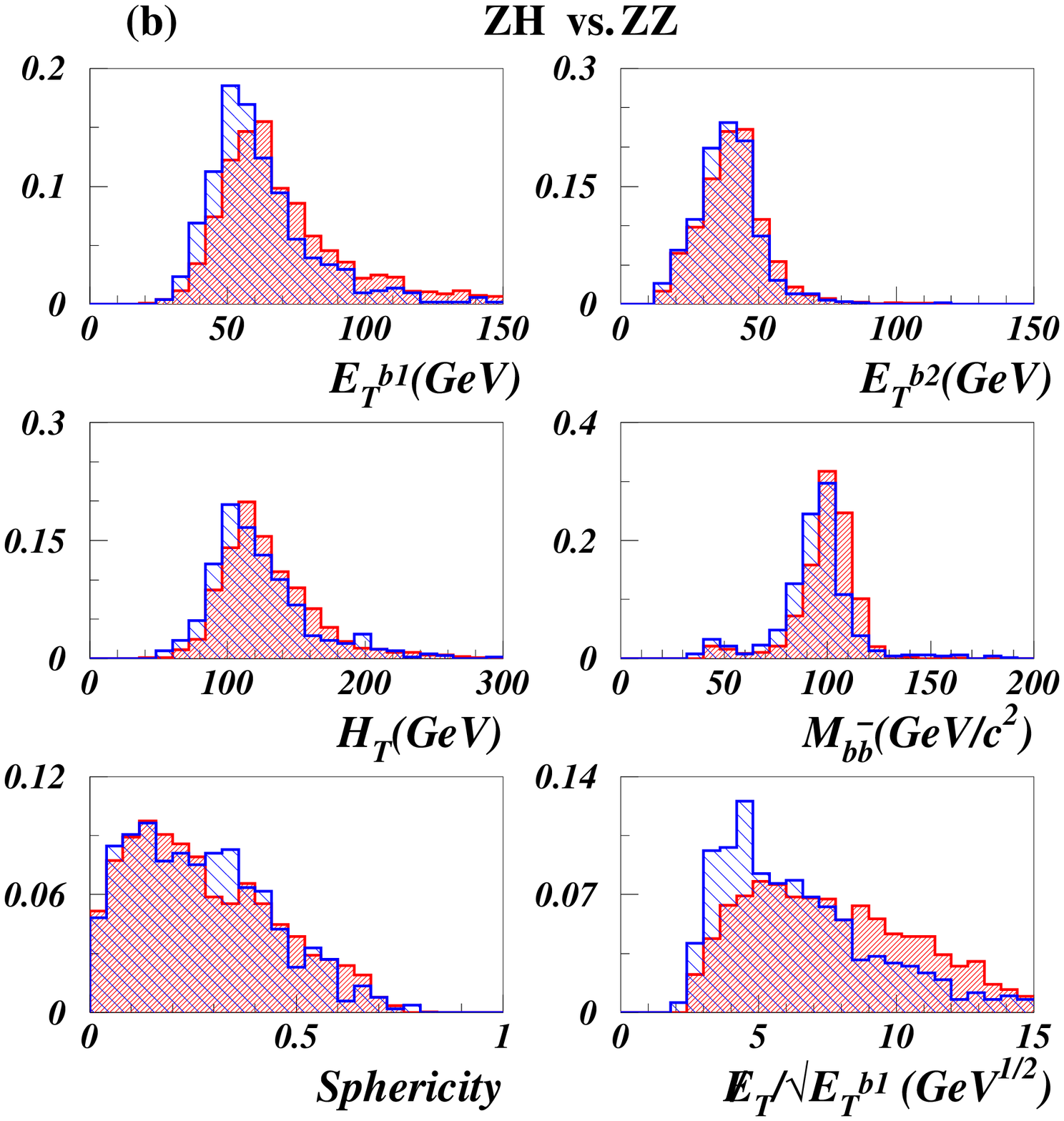, width=3.2in}}
    \end{minipage}
    \begin{minipage}[b]{.45\linewidth}
    \mbox{\epsfig{file=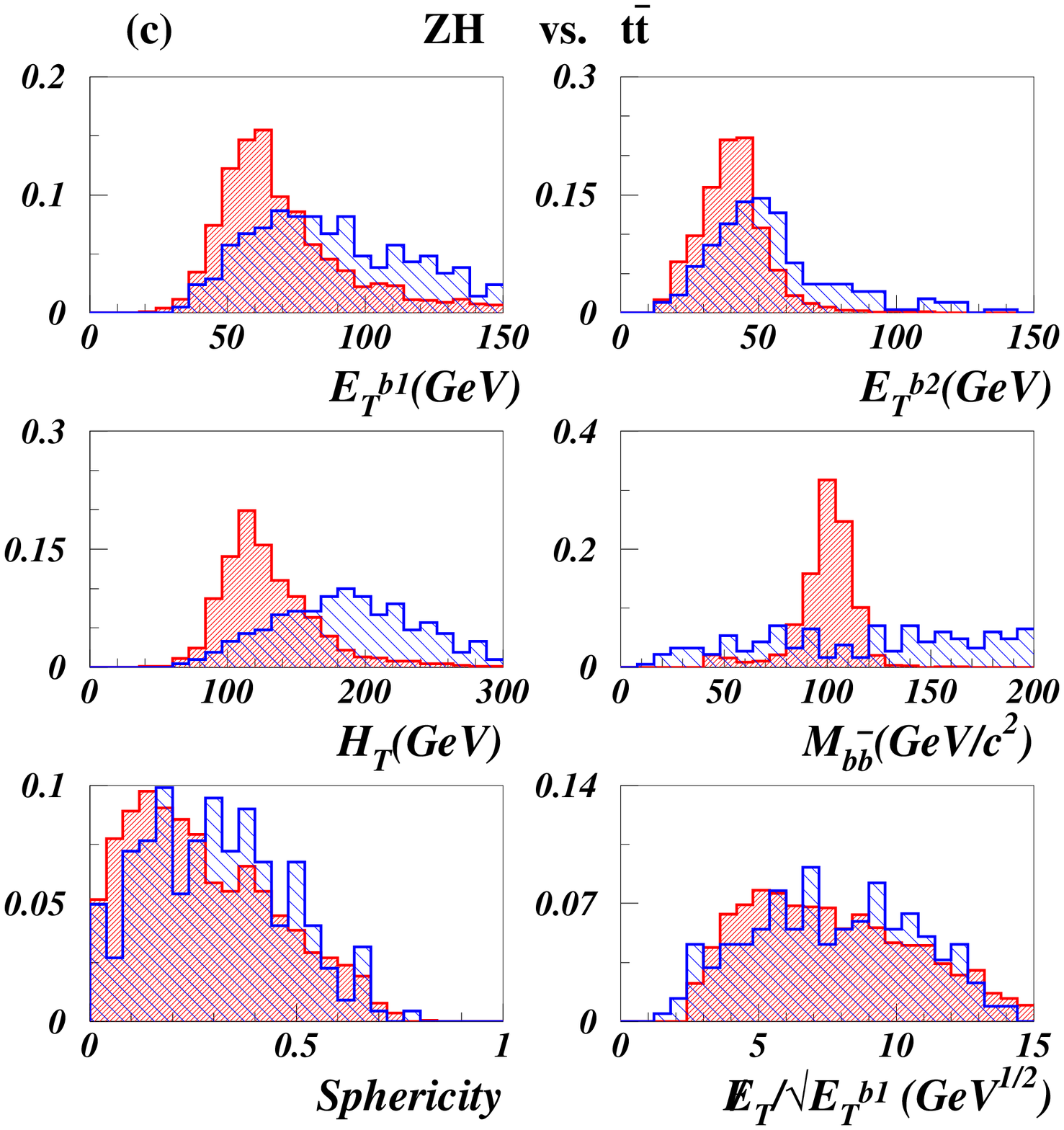, width=3.2in}}
    \end{minipage}
    \begin{minipage}[b]{.45\linewidth}
    \mbox{\epsfig{file=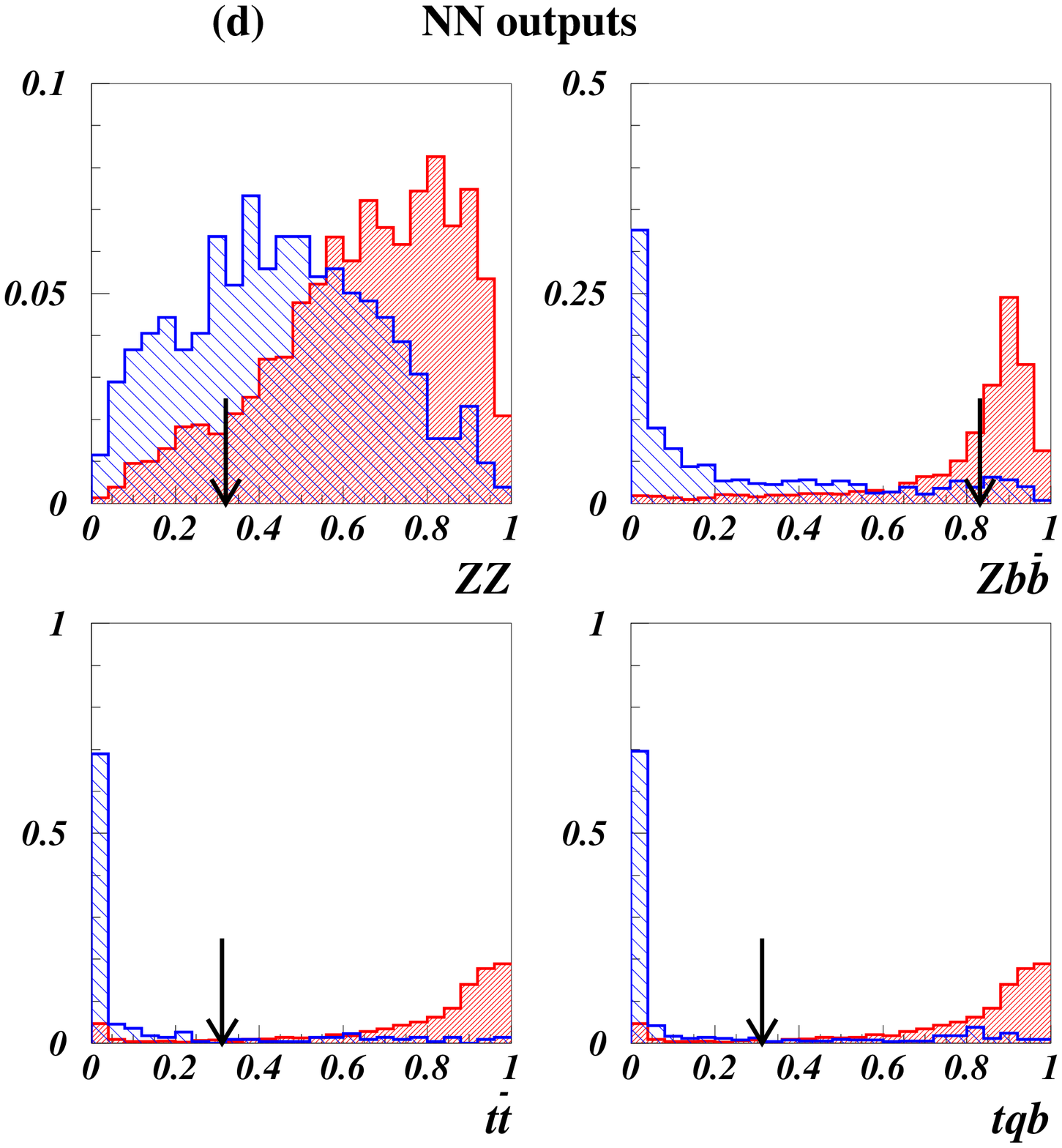, width=3.2in}}
    \end{minipage}
\vspace{0.3in}

\caption[]{Missing transverse energy channel.
Distributions of variables used in training the neural networks
for signal (with $M_H$ = 100~\gevcc) and different backgrounds,
together with distributions of network outputs. (a) Signal \vs\ $Z
b\bar{b}$; (b) signal \vs\ $ZZ$; (c) signal \vs\ $t\bar{t}$ and (d)
distributions of neural network outputs for networks trained using
signal \vs\ the backgrounds $ZZ$, $Zb\bar{b}$ and $t\bar{t}$.
The signal histograms are heavily shaded. 
The arrows indicate the cuts.
}

\label{fig:zhnn}

\end{figure}

The centrality, ${\cal C}$, has larger mean value
(as is the case with $S$)
for signal events than for backgrounds. 
The variable $\frac{\met}{\sqrt{E^{b1}_T}}$ is a measure of the significance
of the missing transverse energy.
The smallest of azimuthal angles between \met and the jets in the event is
expected to be smaller for  $Wb\bar{b}$,
$Zb\bar{b}$ as well as  high multiplicity  $t\bar{t}$ events 
than in signal events.
We show the
distributions of the variables and neural
network outputs in Figs.~\ref{fig:zhnn}(a-d).
Again the signal distributions are for $M_H$=100 \gevcc.
The results for this channel, after optimized cuts on network outputs,
are listed in Table V.

\section{Discussion and Summary}
\label{sec:summary}

\begin{figure}
    \addfig{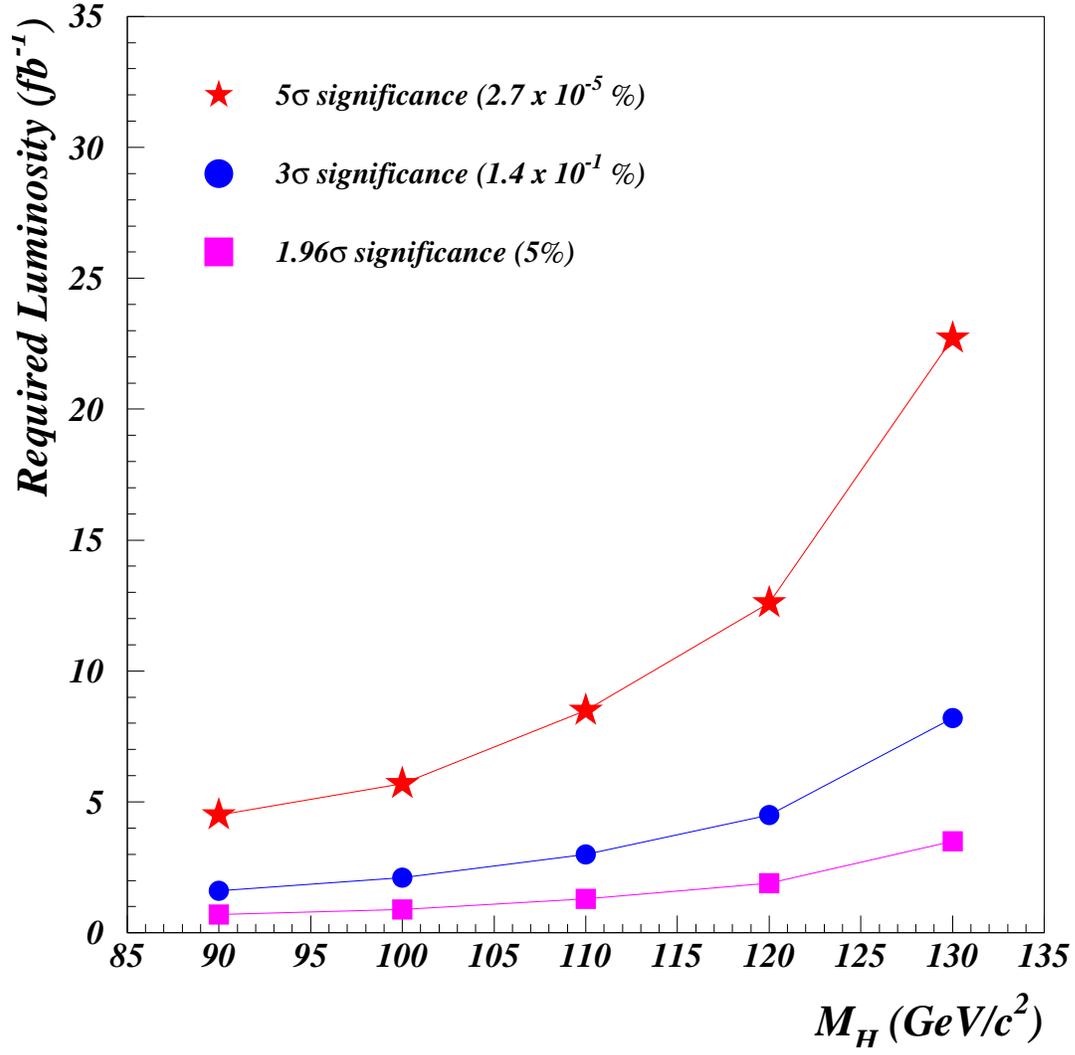}{6.0in}
	\vspace{0.3in}
    \caption[]{Required integrated luminosity, with all channels
combined, at
    $5\sigma$, $3\sigma$ and $1.96\sigma$ (95\% C. L.) significance, for
NN analysis.
}
    \label{fig:lumcomb}
\end{figure}

\begin{figure}
    \addfig{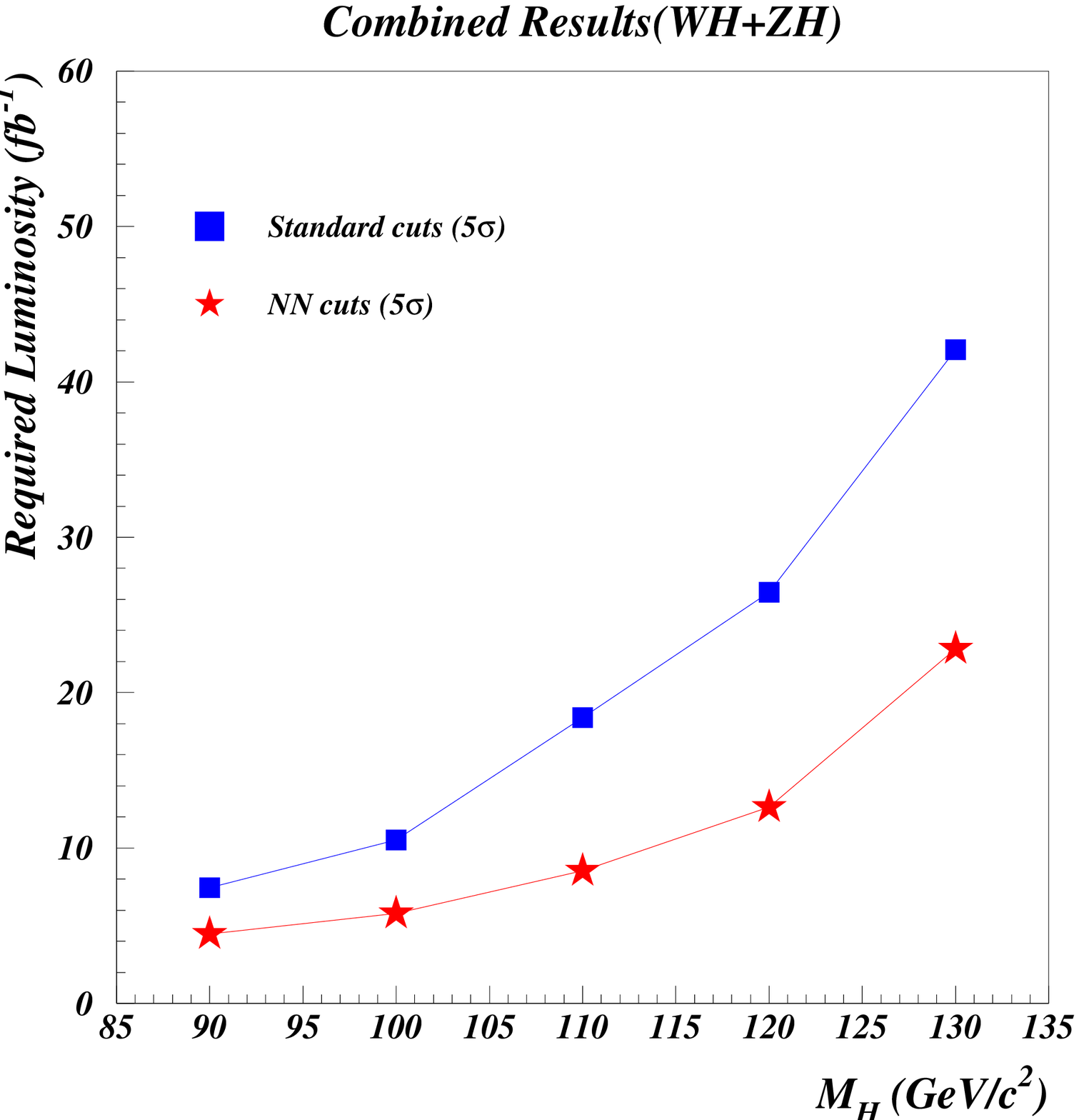}{6.0in}
\vspace{0.3in}
    \caption{
    Comparison of required integrated luminosity for a $5\sigma$ observation
    with all channels combined for NN and standard cuts.
    The luminosities given are for a {\em  single} Tevatron experiment, as
in the previous plots. For a given integrated luminosity the NN analysis
provides a much higher discovery reach in mass.}
\label{fig:whzhcomb}
\end{figure}

In Table VI we compare the results of our
multivariate analysis with those based on the standard cuts, while
Table VII and Figs.~\ref{fig:lumcomb} and \ref{fig:whzhcomb} 
show our final results,
where we have combined
all channels. The striking feature of these results is the
substantial reduction in integrated luminosity
required to make a $5\sigma$ discovery of
the Higgs boson if one adopts a multivariate approach instead of
the traditional method based on univariate cuts.
In each of the
three channels, the signal significance, 
which we define as $S/\sqrt{B}$, is seen to
be 20-60\% 
higher from our multivariate analysis as compared to an optimal
conventional analysis.
 For example, at $M_H = 110$ \gevcc\
we find that the required integrated luminosity for 
a $5\sigma$ observation  decreases from 18.3 fb$^{-1}$
to 8.5 fb$^{-1}$. 
The results in Table VII include statistical errors only. 
The dominant systematic error will likely be
due to background modeling. However, given the large data-sets expected
by the end of Run II we can anticipate that
a thorough experimental study of the
relevant backgrounds will have been undertaken. Therefore, it is
possible that systematic errors could, eventually, be reduced
to well under $10$\%. We can estimate the effect
of systematic error by adding it in quadrature 
to the statistical error. If we assume a 10\% 
systematic error on the total background the required integrated
luminosity for a $5\sigma$ observation increases from 8.5 fb$^{-1}$
to 12.8 fb$^{-1}$.

Run II at the Tevatron with the CDF and D\O\ detectors
will begin in early 2001. Recently the scope of Run II has been
expanded. The goal (hope) is to collect about 15-20 fb$^{-1}$ per
experiment in the period up to and including the start of the LHC.
After 5 years
of running, each experiment could see a 3$\sigma$-5$\sigma$ signal of
a neutral Higgs boson with $M_H \leq$ 130 \gevcc.
This exciting possibility for the
Tevatron is the principal motivation for the recent
important decision to expand the scope of
Run II in order to accumulate as much data as possible. However, even with
the expanded scope a discovery may be possible only if these
data are analyzed with the most efficient
methods available, such as the one we have described in this paper.
It is important to note that the results we have presented are for a
{\em single} experiment. That is, our conclusion is that each experiment
has the potential of making an independent discovery.  If the
experiments combine their results the discovery of a low-mass
Higgs boson at the Tevatron might be at hand a lot sooner.

\acknowledgements
We thank the members of the Run II Higgs Working Group
and, in particular, Ela Barberis, Alexander Belyaev, John Conway, John Hobbs,
Rick Jesik, Maria Roco and Weiming Yao
for useful discussions and for help with the
event simulation. The research was supported in part by the U.S. Department of
Energy under contract numbers DE-AC02-76CHO3000 and DE-FG02-97ER41022.
This work was carried out by the authors as part of the Higgs Working
Group\footnote{Run II Higgs Working Group (Run II SUSY/Higgs
workshop).\\
http://fnth37.fnal.gov/higgs.html.} study at Fermilab.


\begin{table}
\begin{center}
\begin{tabular}{|lc|lc|lc|}
\multicolumn{2}{c} {$WH\rightarrow \ell\nu b\bar{b}$}  &
\multicolumn{2}{c} {$ZH \rightarrow$ $\ell^{+}\ell^{-}$
$b\bar{b}$}  & \multicolumn{2}{c} {$ZH \rightarrow \nu \bar{\nu}
b\bar{b}$}\\ \hline

$M_H$ (GeV/c$^2$) & $\sigma\times BR(\mbox{fb})$  & $M_H$
(GeV/c$^2$) & $\sigma\times BR(\mbox{fb})$  & $M_H$ (GeV/c$^2$) &
$\sigma\times BR(\mbox{fb})$  \\ \hline

90      & 119.0 & 90    & 20.3  & 90     & 40.6 \\ 100     & 85.4
& 100   & 14.8  & 100    & 29.6 \\ 110     &  62.3 & 110   & 10.9
& 110    & 21.8 \\ 120     &  45.3 & 120   & 8.22  & 120    & 16.4
\\ 130     & 34.1  & 130   & 6.25  & 130    & 12.5 \\ \hline

Backgrounds & & & & &   \\ $Wb\bar{b}$ &  3500.0   & $Zb\bar{b}$
&  350.0    &  $Zb\bar{b}$  &  700.0    \\ $WZ$        &   164.8 &
& & &   \\ $tbq$       & 800.0 & $tbq$ &  800.0 & $tbq$   &
800.0\\ \hline
            & $\sigma$ (fb)   &        & $\sigma$ (fb) &  & $\sigma$ (fb)\\
            &         &  $ZZ$ & 1235.0 & $ZZ$    & 1235.0  \\
$tb$        &  1000.0 &  $tb$ & 1000.0 & $tb$    & 1000.0  \\
$t\bar{t}$  &  7500.0 &  $t\bar{t}$    &  7500.0 & $t\bar{t}$ &
7500.0 \\
\end{tabular}

\caption[]{Cross section times branching ratio for the 
$WH$ and $ZH$ processes we have studied, for various
$M_H\cite{spira}$ and for the various backgrounds. 
Note: For $tb$, $t\bar{t}$ and $ZZ$ processes
we give the total cross section.}
\end{center}
\label{tab:xsections}
\end{table}

\begin{table}[h!]
\label{tab:whvariables}
\begin{center}
\begin{tabular}{|ccc|}
$Wb\bar{b}$         & $WZ$ & $t\bar{t}$\\ \hline $E_T^{b1}$  &
$E_T^{b1}$  & $E_T^{b1}$\\ $E_T^{b2}$  & $E_T^{b2}$  &
$E_T^{b2}$\\ $M_{b\bar{b}}$ & $M_{b\bar{b}}$ & $M_{b\bar{b}}$\\
$H_T$       & $H_T$       & $H_T$\\ $E_T^{\ell}$ & $E_T^{\ell}$ &
$\met$\\ $S$  & $S$  & $\Delta R(b_1,\ell)$\\ $\Delta R(b_1,b_2)$
& $\eta_{\ell}$ & $\Delta R(b_1,b_2)$\\
\end{tabular}
\caption[]{Single lepton channel. 
Variables used in training the neural networks for
signals against specific backgrounds.}
\end{center}
\end{table}

\begin{table}[h!]
\label{tab:whresults}
\begin{center}
\begin{tabular}{lrrrrr}
$M_H$~GeV/c$^2$    &90 &100    &110    &120    &130
 \\\hline

Number of events(1 fb$^{-1}$)     &       &       &       &       &
\\
 $WH$               &8.65   &8.97   &4.81   &4.41   &3.71   \\ \hline
 $Wb\bar{b}$        &12.28  &12.48  &5.84   &9.66   &20.12  \\
 $WZ$               &7.52   &10.32  &1.72   &1.00   &0.97   \\
 $tqb$              &0.51   &0.95   &0.58   &0.71   &0.96   \\
 $tb$               &2.46   &5.40   &3.44   &5.89   &9.33   \\
 $t\bar{t}$         &5.63   &9.89   &7.24   &8.39   &14.49  \\ \hline

 Total background   &28.40  &39.04  &18.81  &25.67  &45.87  \\\hline
 Signal significance&       &       &       &       &       \\
 S/B                &0.31   &0.23   &0.26   &0.17   &0.081  \\
 S/$\sqrt{B}$ (1 fb$^{-1}$)
                    &  1.62 &  1.44 &  1.11 &  0.87 &  0.55 \\
 S/$\sqrt{B}$ (2 fb$^{-1}$)
                    &  2.29 &  2.04 &  1.57 &  1.23 &  0.78 \\
 S/$\sqrt{B}$ (30 fb$^{-1}$)
                    &8.87   &  7.89 &  6.08 &  4.77 &  3.01 \\\hline
Required luminosity (fb$^{-1}$)
                    &       &       &       &       &       \\
$5\sigma$           & 9.5   & 12.1  & 20.3  & 33.0  & 82.6  \\
$3\sigma$           & 3.4   &  4.3  &  7.3  & 11.9  &  29.8 \\
$1.96\sigma$ (95\% CL)
                    & 1.5   &  1.9  &  3.1  &  5.1  &  12.7 \\
\end{tabular}

\caption[]{Single lepton channel. Results for the number of signal and
background events (top portion of the table) for 1 fb$^{-1}$
of integrated luminosity. The cuts on the network outputs were chosen
to yield maximum significance for each Higgs boson mass, leading to
different background counts at each mass. }
\end{center}
\end{table}

\begin{table}[h!]
\label{tab:zhllresults}
\begin{center}
\begin{tabular}{lrrrrr}
$M_H$ (GeV/c$^2$)   &           90 &   100 &   110 &   120 &
130\\ \hline Number of events &&&&& \\ 
$ZH$        &   1.26 & 0.87 &   0.79 &   0.80 &   0.58\\ \hline 
$Zb\bar{b}$ &   0.61 &
0.45 &   0.61 &   1.50 &   1.42\\ 
$ZZ$        &   2.04 &   1.44 & 1.42 &   0.83 &   0.31\\ 
$t\bar{t}$  &   0.28 &   0.05 &   0.23 & 0.44 &   0.18\\ \hline 
Total background & 2.93 & 1.94 &  2.26 & 2.77 &   1.91\\ \hline 
$S/B$        &   0.43 & 0.45 & 0.35 &  0.29 &  0.31\\ 
$S/\sqrt{B}$ &   0.74 & 0.63 & 0.54 &  0.48 &  0.42\\
\end{tabular}

\caption{Di-lepton channel. Results for 1 fb$^{-1}$.}
\end{center}
\end{table}

\begin{table}[h!]
\label{tab:zhnnresults}
\begin{center}
\begin{tabular}{lrrrrr}
$M_H$ (GeV/c$^2$) &        90 &    100 &   110 &   120 &   130\\
\hline Number of events &&&&& \\ $ZH$ &      6.66 &    4.37 &
3.53 &   2.76 &   2.16\\ $WH$ &      5.59 &    3.75 &   2.79 &
1.98 &   1.70\\ \hline Total signal&  12.25 &    8.12 &   6.32 &
4.74 &   3.86\\ \hline $Zb\bar{b}$ &   8.12 &    4.97 &   4.83 &
3.85 &   3.92\\ $Wbb$   &      21.70 &   13.12 &  10.68 &   8.22 &
7.53\\ $ZZ$ &         11.24 &    6.14 &   2.59 &   1.05 &   0.59\\
$WZ$ &          7.95 &    4.49 &   1.99 &   0.90 &   0.54\\ $tqb$
&         0.63 &    0.27 &   0.37 &   0.24 &   0.29\\ $tb$ &
6.83 &    2.99 &   4.27 &   5.12 &   6.40\\ $t\bar{t}$ &    5.10 &
2.70 &   3.00 &   3.00 &   4.35\\ \hline Total background & 61.57
&34.8 & 27.73 &  22.38 &  23.62\\ \hline $S/B$ &        0.20 &
0.23 &  0.23 &  0.21 &  0.16\\ $S/\sqrt{B}$ & 1.56 &   1.38 &
1.20 &  1.00 &  0.79\\
\end{tabular}
\caption{Missing transverse energy channel. 
Results for 1 fb$^{-1}$.}
\end{center}
\end{table}

\begin{table}[h!]
\label{tab:comparison}
\begin{center}
\begin{tabular}{lrrrr}
 channel& mass& standard& neural& $L^{NN}$/$L^{std}$\\
 &        (GeV)& cuts& net& (for $5\sigma$ obsv.)\\ \hline
$\ell+ \met + b \bar{b}$& 100& 0.98& 1.44& 0.46\\ & 110&
0.69& 1.11& 0.39\\ & 120& 0.58& 0.87& 0.44\\ & 130& 0.44& 0.55&
0.64\\ \hline 
$\met + b \bar{b}$& 100& 1.09&
1.38& 0.62\\ & 110& 0.85& 1.20& 0.50\\ & 120& 0.67& 1.00& 0.49\\ &
130& 0.54& 0.78& 0.47\\ \hline 
$\ell^+ \ell^- b\bar{b}$& 100& 0.48& 0.63& 0.58\\ 
& 110& 0.40& 0.52& 0.59\\ & 120&
0.40& 0.48& 0.69\\ & 130& 0.33& 0.42& 0.61\\ \hline
\end{tabular}

\caption{ Comparison of $S/\sqrt{B}$ achievable with conventional
and neural networks cuts. Shown in the last column are the ratios of integrated
luminosity required in the multivariate analysis to that required
in the conventional analysis for a $5\sigma$ observation. }
\end{center}
\end{table}

\begin{table}[h!]
\label{tab:combined}
\begin{center}
\begin{tabular}{lrrrrr}
$M_H$ (GeV/c$^2$)         &    90 &  100 &  110 &  120 &  130\\
\hline $S/\sqrt{B}$ (1 fb$^{-1}$)  &   2.4 &  2.1 &  1.7 &  1.4 &
1.0\\ $S/\sqrt{B}$ (2 fb$^{-1}$) &   3.3 &  3.0 &  2.4 &  2.0 &
1.5\\ $S/\sqrt{B}$ (30 fb$^{-1}$) &  12.9 & 11.5 &  9.4 &  7.7 &
5.7\\ \hline Required luminosity &&&&& \\ $5\sigma$ (Conventional)
&   7.5 & 10.5 & 18.3 & 26.6 & 42.2\\ $5\sigma$ (NN)            &
4.5 &  5.7 &  8.5 & 12.6 & 22.7\\ $3\sigma$ (NN)            &
1.6 &  2.1 &  3.0 &  4.5 &  8.2\\ 95\% CL   (NN)            &
0.7 &  0.9 &  1.3 &  1.9 &  3.5\\
\end{tabular}
\caption{Combined results of all three channels.  We have
simply added the signal counts and background counts from all
three channels to get the total expected signal and background,
respectively. }
\end{center}
\end{table}


\begin{references}

\bibitem{smreview}
J.~Erler and P.~Langacker, in 
{\em Proceedings of the 5th International WEIN Symposium: 
A Conference on Physics Beyond the Standard Model,} Santa Fe, NM, 1998;
e-print hep-ph/9809352.

\bibitem{altarelli}
G.~Altarelli, CERN report, CERN-TH/97-278; e-print hep-ph/9710434.

\bibitem{topreview}
P.~C.~Bhat, H.~B.~Prosper and S.~S.~Snyder, Int. J. Mod. Phys. 
{\bf A13}, 5113 (1998).

\bibitem{lepewwg}
LEP Electroweak Group, http://www.cern.ch/LEPEWWG/plots/summer99.

\bibitem{martin}
See for example, S.~Martin, in {\em Perspectives on Supersymmetry,} edited by
G.~L.~Kane (World Scientific, Singapore, 1998);
e-print hep-ph/9709356 v3, 1999.

\bibitem{carena}
M.~Carena, M.~Quiros, C.~E.~M.~Wagner, Nucl. Phys. {\bf B}461, 407 (1996).

\bibitem{ukreport}
B.C.~Alanach, {\em et al.}, { Report of the Beyond the Standard
Model Working Group of the 1999 UK Phenomenology Workshop on
Collider Physics (Durham)}; e-print hep-ph/9912302.

\bibitem{workshop}
Run II Higgs Working Group of the Run II SUSY/Higgs workshop.\\
http://fnth37.fnal.gov/higgs.html;
T.~Han, A.~S.~Turcot, and R.~Zhang,
\Journal{\PRD}{59}{093001}{1999}.

\bibitem{lepc}
A.~Sopczak, e-print hep-ph/0004015, IEKP-KA/2000-06.\\
See also,
http://l3www.cern.ch/conferences/talks99.html.

\bibitem{spira}
M.~Spira, e-print hep-ph/9810289, A.~Djouadi, J.~Kalinowski, and M.~Spira, 
e-print hep-ph/9808312.


\bibitem{bhat}
P.C. Bhat (for the D\O\ collaboration), in {\em Proceedings of the 10th
Topical Workshop on proton-antiproton Collider Physics,} Batavia, IL
(AIP, Woodbury, NY, 1995), p. 308;
C.~M.~Bishop, {\em Neural Networks for Pattern Recognition,}
(Clarendon Press, Oxford, 1998);
R.~Beale and T.~Jackson, {\it Neural Computing: An Introduction,}
(Adam Hilger, New York, 1991).

\bibitem{blum}
D.W.~Ruck {\em et al.}, IEEE Trans. Neural Networks
{\bf 1 (4)}, 296 (1990);
E.A.~Wan, 
IEEE Trans. {\em ibid.} {\bf 1 (4)}, 303 (1990);
E.K.~Blum and L.K.~Li, 
Neural Networks, {\bf 4}, 511 (1991).


\bibitem {jetnet}
JETNET, C.~Peterson, J.~R\"{o}gnvaldsson, and L.~L\"{o}nnblad, 
Comput. Phys. Commun.  {\bf 81}, 185 (1994).
We used JETNET version 3.0.

\bibitem{pythia}
PYTHIA,
T. Sj\"{o}strand, Comput. Phys. Commun. {\bf 82}, 74 (1994).

\bibitem{comphep}
CompHEP,
A.~S.~Belyaev, A.V.~Gladyshev and A.V.~Semenov,
e-print  hep-ph/9712303; E.E.~Boos 
\etal, e-print hep-ph/9503280.


\bibitem{SHW}
SHW 2.0, 
J.~Conway, available at\\ 
http://www.physics.rutgers.edu/$\sim$jconway/soft/shw/shw.html
(unpublished).


\bibitem{rgs}
H.~B.~Prosper \etal, (for the D\O\ Collaboration), 
in {\em Proceedings of the  International Conference
on Computing in High Energy Physics '95} (Rio de Janeiro, Brazil)
(World Scientific, River Edge, NJ, 1996).


\end{references}
\end{document}